\documentclass[prl,twocolumn]{revtex4}  

\usepackage{amssymb}
\usepackage{amsmath}
\usepackage{graphicx}
\usepackage{bbm}
\usepackage{color}
\usepackage{comment}
\usepackage{gensymb}
\definecolor{blue}{rgb}{0,0,1}
\definecolor{grey}{rgb}{0.6,0.6,0.6}

\newcommand{\bra}[1]{\langle #1 |}
\newcommand{\ket}[1]{| #1 \rangle}

\begin{document}

\title{Phonon networks with SiV centers in diamond waveguides}
\author{M.-A. Lemonde$^1$, S. Meesala$^2$, A. Sipahigil$^{3,4}$, M. J. A. Schuetz$^4$, M. D. Lukin$^4$, M. Loncar$^2$, P. Rabl$^1$}
\affiliation{$^1$ Vienna Center for Quantum Science and Technology,
Atominstitut, TU Wien, 1040 Vienna, Austria}
\affiliation{$^2$ John A. Paulson School of Engineering and Applied Sciences, Harvard University, 29 Oxford Street, Cambridge, MA 02138, USA}
\affiliation{$^3$  Institute for Quantum Information and Matter and Thomas J. Watson, Sr., Laboratory of Applied Physics, California Institute of Technology, Pasadena, California 91125, USA}
\affiliation{$^4$  Department of Physics, Harvard University, Cambridge, Massachusetts 02138, USA}

\date{\today}

\begin{abstract}
We propose and analyze a novel realization of a solid-state quantum network, where separated silicon-vacancy centers are coupled via the phonon modes of a quasi-1D diamond waveguide. In our approach, quantum states encoded in long-lived electronic spin states can be converted into propagating phonon wavepackets and be  reabsorbed efficiently by a distant defect center. Our analysis shows that under realistic conditions, this approach enables the implementation of high-fidelity, scalable quantum communication protocols within chip-scale spin-qubit networks. Apart from quantum information processing, this setup constitutes a novel waveguide QED platform, where strong-coupling effects between solid-state defects and individual propagating phonons can be explored at the quantum level. 
\end{abstract}

\maketitle

 
Electronic and nuclear spins associated with defects in solids comprise a promising platform for the realization of practical quantum technologies~\cite{Ladd2010}. 
A prominent example is the nitrogen-vacancy (NV) center in diamond~\cite{Doherty2013,Childress2014}, for which techniques for state detection~\cite{Jelezko2004}, coherent manipulations~\cite{Childress2006,Lange2010,Bassett2014} and local entanglement operations~\cite{Pfaff2013, Dolde2013,Waldherr2014} have been demonstrated and employed, for example, for various nanoscale sensing applications~\cite{Rondin2014}. Despite this progress in the local control of spin qubits, integrating many spins into larger networks remains a challenging task. To achieve this goal, several schemes for interfacing spins via mechanical degrees of freedom have recently been discussed~\cite{Rabl2010,Habraken2012, Bennett2013, Albrecht2013,Schuetz2015,Lee2017} and first experiments demonstrating magnetic~\cite{Rugar2004, Arcizet2011,Kolkowitz2012} or strain-induced~\cite{MacQuarrie2013,Ovartchaiyapong2014,Barfuss2015,Meesala2016,Golter2016} couplings of mechanical vibrations to both long-lived spin states and electronic excited states of NV centers have been carried out. However, the weak intrinsic coupling of spins to vibrational modes and the short coherence of optically excited states make the extension of these methods into the quantum regime challenging.

In this Letter we describe  the implementation of a phonon quantum network, where negatively-charged silicon-vacancy (SiV) centers are coupled via  propagating phonon modes of a 1D diamond waveguide~\cite{Burek2012,Khanaliloo2015,Sipahigil2016,Mouradian2017}. 
The electronic ground state of the SiV center features both spin and orbital degrees of freedom~\cite{Goss1996,Hepp2014,HeppThesis}, which makes it naturally suited for this task; quantum states can be encoded in long-lived superpositions of the two lowest spin-orbit-coupled states~\cite{Rogers2014, Becker2016,Zhou2017,Sukachev2017,Becker2017}, while a controlled admixing of higher orbital states, which are susceptible to strain,  gives rise to a strong and tunable coupling to phonons. The central phonon frequency of $\sim 46$ GHz set by the large spin-orbit splitting enables quantum-coherent operations already at convenient temperatures of $T\lesssim 1$ K, when thermal excitations at this frequency are frozen out.
Our analysis shows that high-fidelity quantum state transfer protocols between distant SiV centers can be implemented under realistic conditions. Moreover, we propose a scalable operation of such phonon networks using switchable single-defect mirrors. 


\begin{figure}[t]
	\begin{center}
	\includegraphics[width= 0.9\columnwidth]{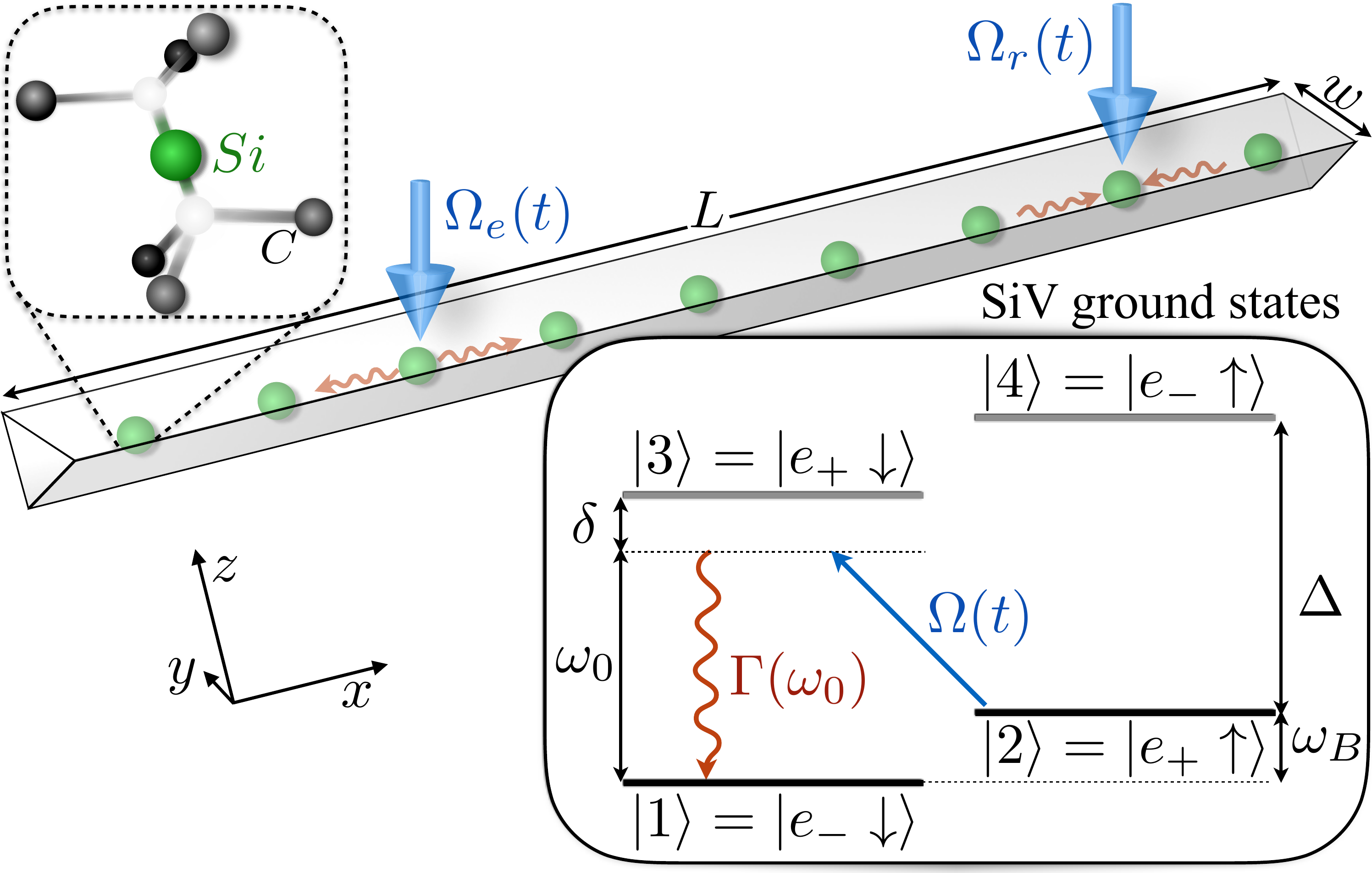}
	\end{center}
	\vspace{-0.5cm} 
	\caption {Setup. An array of SiV defects is embedded in a 1D phonon waveguide. The inset shows the level structure of the electronic ground state of the SiV center. A tunable Raman process involving the excited state $|3\rangle$ is used to coherently convert the population of the stable state $|2\rangle$ into a propagating phonon, which can be reabsorbed by any other selected center along the waveguide. 
See text for more details.
}
\label{Fig:Schema}
\end{figure}


\textit{Model.}---We consider a system as depicted in Fig.~\ref{Fig:Schema}, where an array of SiV centers is embedded in a 1D diamond waveguide. The electronic ground state of the SiV center is formed by an unpaired hole of spin $S=1/2$, which occupies one of the two degenerate orbital states $|e_x\rangle$ and $|e_y\rangle$. In the presence of spin-orbit interactions and a weak Jahn-Teller effect, the four states are split into two doublets, $\{\ket1\simeq  \ket{e_-,\downarrow}, \ket2\simeq |e_+,\uparrow\rangle\}$ and $\{\ket3\simeq|e_+,\downarrow\rangle, \ket4\simeq|e_-,\uparrow\rangle\}$, which are separated by $\Delta/2\pi \simeq 46$ GHz~\cite{Hepp2014, HeppThesis}. Here, $\ket{e_{\pm}} =  (\ket{e_{x}}\pm i\ket{e_y})/\sqrt{2}$ are eigenstates of the orbital angular momentum operator, i.e.~$L_z\ket{e_{\pm}} = \pm\hbar\ket{e_{\pm}}$, where the $z$-axis is along the symmetry axis of the defect. 
In the presence of a magnetic field $\vec B=B_0\vec e_z$, the Hamilton operator for a single SiV center is $(\hbar=1)$
\begin{equation}\label{Eq:HSiV}
\begin{split}
H_{\rm SiV}  &= \omega_B  \ket2 \bra2 + \Delta\ket3 \bra3 + (\Delta + \omega_B)\ket4 \bra4\\
	&+\frac{1}{2}\left[\Omega(t)e^{i[\omega_{d} t + \theta(t)]} \left(\ket2 \bra3 +  \ket1 \bra4  \right) +{\rm H.c.}\right],
	\end{split}
\end{equation}
where $\omega_B=\gamma_sB_0$  and  $\gamma_s$ is the spin gyromagnetic ratio. 
In Eq.~\eqref{Eq:HSiV}, we have included a time-dependent driving field with a tunable Rabi-frequency $\Omega(t)$ and phase $\theta(t)$, which couples the lower and upper states of opposite spin. This coupling can be implemented either directly with a microwave field of frequency $\omega_{d}\sim\Delta$~\cite{Pla2012}, or indirectly via an equivalent optical Raman process~\cite{SI}. The latter method is already used in experiments to initialize and prepare individual SiV centers in superpositions of $|1\rangle$ and $|2\rangle$~\cite{Rogers2014, Becker2016, Zhou2017} with coherence times that can exceed 10~ms in the absence of thermal processes and with dynamical decoupling~\cite{Sukachev2017}.
Further details on the derivation of $H_{\rm SiV}$ are given in the supplementary material~\cite{SI}.

For the waveguide, we consider a quasi-1D geometry
of width $w$ and length $L\gg w$. 
The waveguide supports travelling phonon modes of frequency $\omega_{n,k}$ and mode function $\vec u_{n,k}(\vec r)\sim\vec u^\perp_{n,k}(y,z)e^{ikx}$, where $k$ is the wavevector along the waveguide direction, $n$ is the branch index and $\vec u^\perp_{n,k}(y,z)$ is the transverse profile of the  displacement field. 
The phonons induce transitions between the orbital states $\ket{e_\pm}$~\cite{Sohn2017,Jahnke2015, Kepesidis2016}, and the Hamiltonian for the whole system reads
\begin{equation}
\begin{split}
H&= \sum_j H_{\rm SiV}^{(j)} +\sum_{n,k} \omega_{n,k} a_{n,k}^\dag  a_{n,k}\\
&+\frac{1}{\sqrt{L}}\sum_{j,k,n}   \,  \left( g^j_{n,k}J^j_{+} a_{n,k} e^{ikx_j}   + {\rm H.c.}
\right).
\end{split}
\label{Eq:Hnetwork}
\end{equation}
Here $j$ labels the SiV centers located at positions $\vec r_j=(x_j,y_j,z_j)$, $J_- = (J_+)^\dag = \ket1\bra3 + \ket2\bra4$ is the spin-conserving lowering operator and $a_{n,k}$ ($a_{n,k}^\dag$) are the annihilation (creation) operators for the phonon modes. The couplings $g^j_{n,k}\equiv g_{n,k}(y_j,z_j)$ depend on the components of the local strain tensor, $\epsilon_{n,k}^{ab}(\vec r_j) = \frac{1}{2}[ \frac{\partial }{\partial x_b} u^a_{n,k}(\vec r_j) + \frac{\partial }{\partial x_a}  u^b_{n,k}(\vec r_j)]$, and can be evaluated for a known transverse mode profile $\vec u^\perp_n(y,z)$~\cite{SI, Kepesidis2016}. We express the resulting couplings as
\begin{equation}
	g^j_{n, k} = d \sqrt{\frac{\hbar k^2}{2\rho A\omega_{n,k}}} \xi_{n,k} (y_j,z_j),
	\label{Eq:gstrain}
\end{equation}
where $d/2\pi\sim 1$ PHz is the strain sensitivity of the orbital states~\cite{Jahnke2015, Sohn2017}, $\rho$ the density and $A$ the transverse area of the waveguide. The dimensionless coupling profile $\xi_{n,k}(y,z)$ accounts for the specific strain distribution  and $\xi (y,z)=1$ for a homogeneous compression mode.


\begin{figure}
	\begin{center}
	\includegraphics[width=0.9\columnwidth]{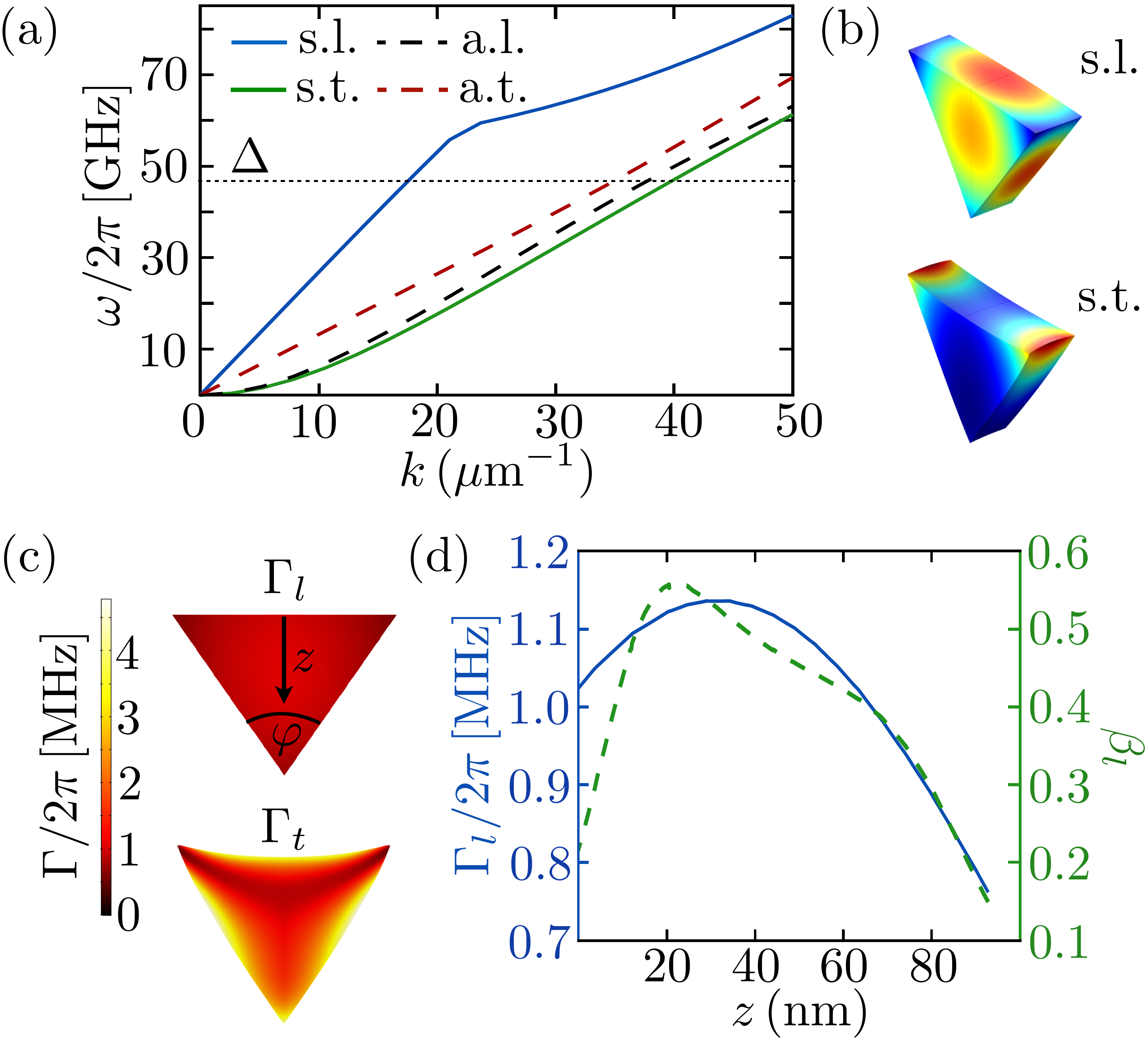}
	\end{center}
	\vspace{-0.5cm} 
	\caption{Phonon waveguide. (a) Acoustic dispersion relation for a triangular waveguide of width $w = 130$ nm and etch-angle $\varphi = 35^o$. Symmetric (solid lines) and anti-symmetric  (dashed lines) branches with respect to the vertical mirror-symmetry plane are shown.
(b) Normalized displacement profiles of the symmetric phonons at 46 GHz. 
	(c) The emission rates into the symmetric longitudinal ($\Gamma_{l}$) and transverse ($\Gamma_{t}$) polarization modes at 46 GHz are plotted for different positions of the  SiV center within the triangular cross-section.
	(d) $\Gamma_l$ and fraction ($\beta_l$) of spontaneous emission into the longitudinal branch for different positions of the SiV center along the vertical mirror-symmetry axis. For all results, an orientation of the waveguide along the [110] crystal axis of diamond and SiV centers oriented along [$\bar{1}$11] and [1$\bar{1}$1], i.e., orthogonal to the waveguide axis, have been assumed.}
\label{Fig:PhononModes}
\end{figure}


\emph{From cavity to waveguide QED.}---For small structures, $L\sim 10-100\,\mu$m, $w\lesssim 200$ nm, and group velocities $v \sim 10^4$ m/s, the individual phonon modes are well separated in frequency, $\Delta \omega/2\pi\gtrsim 50$ MHz, and the SiV centers can be coupled to a single standing-wave mode with a strength $g_L=g_0\sqrt{\lambda/L}\approx 2\pi\times (4-14)$ MHz, where $g_0/2\pi\approx   105$ MHz and $\lambda\approx 200$ nm is the phonon wavelength.  The system dynamics is then governed  by a Jaynes-Cummings-type interaction between phonons and orbital states~\cite{SI}. In the strong coupling regime, $g_L > \kappa=\Delta/Q$, which is reached for moderate mechanical quality factors of $Q>10^4$, a coherent exchange of phonons and defect excitations becomes possible.  For longer waveguides, the coupling to the quasi-continuum of phonon modes is characterized by the resulting decay rate $\Gamma_j(\Delta)=\sum_n \Gamma_{j,n}(\Delta)$ for  states $|3\rangle$ and $|4\rangle$, where
\begin{equation}
\Gamma_{j,n}(\omega)= \lim_{L\rightarrow\infty} \frac{2\pi}{L}\sum_{k} |g^j_{n,k}|^2 \delta(\omega-\omega_{n,k}). 
\end{equation}  
For a single compression mode with $\vec u^\perp(y,z) \sim \vec x$ and a linear dispersion $\omega_k=v k$, we obtain $\Gamma(\omega) = d^2\hbar\omega/(\rho A v^3)$, which  results in a characteristic phonon emission rate of $\Gamma(\Delta)/2\pi\sim 1$ MHz~\cite{Kepesidis2016}. 

Figure~\ref{Fig:PhononModes} summarizes the simulated  acoustic dispersion relations and the resulting decay rates for a triangular waveguide~\cite{Burek2012,Sohn2017} of width $w=130$ nm. The SiV centers couple primarily to a longitudinal ($l$) compression  and a transverse ($t$) flexural mode with 
group velocities $v_l= 1.71\times10^4$ m/s and $v_t= 0.73 \times 10^4$ m/s, respectively. The coupling to the other two branches of odd symmetry can be neglected for defects near the center of the waveguide. Fig.~\ref{Fig:PhononModes}(c) and (d) show that the rates $\Gamma_{l,t}$ are quite insensitive to the exact location of the SiV center. However, the fraction of phonons emitted into a specific branch, $\beta_n=\Gamma_n/\Gamma$, is significantly below unity as emission is split between a pair of modes. In optical waveguides~\cite{Lodahl2015}, a value of  $\beta<1$ usually arises from the emission of photons into non-guided modes, which are irreversibly lost. 
For a phonon waveguide this is not the case, but the multi-branch nature of the waveguide must be fully taken into account. In all examples below we assume $\beta_l=\beta_t=0.5$, which is most relevant for SiV defects located near the center of the beam.

\emph{Coherent spin-phonon interface.}---We are interested in the transfer of a qubit state, encoded into the stable states $|1\rangle$ and $|2\rangle$, between an arbitrary pair of emitting ($e$) and receiving ($r$) defects in the waveguide, 
\begin{equation}
(\alpha|1\rangle_e+\beta|2\rangle_e)|1\rangle_r \rightarrow |1\rangle_e(\alpha |1\rangle_r +\beta|2\rangle_r  ).
\end{equation}
 As shown in Fig.~\ref{Fig:Schema}, this can be achieved by inducing a 
 Raman transition via state $|3\rangle_e$ to convert the population in state $|2\rangle_e$ into a propagating phonon and by reverting the process at the receiving center. 
 For low enough temperatures, $T\ll \hbar\Delta/k_B\approx 2.2$ K, such that all phonon modes are initially in the vacuum state, this scenario is  described by the following ansatz for the wavefunction  $|\psi(t)\rangle =[\alpha \mathbbm{1} +\beta C^\dag(t)] \ket{\bar 1,0}$, where $\ket{\bar 1,0}$ is the ground state with all SiV centers in state $|1\rangle$ and $C^\dag(t) = \sum_{j=e,r}  \big[ c_j(t)e^{-i\omega_Bt} |2\rangle_j\langle 1|   + b_j(t)e^{-i\omega_0t} |3\rangle_j\langle 1| + \sum_{n,k}  c_{n,k}(t)e^{-i \omega_0t}a^\dag_{n,k}\big]$ creates a single excitation distributed between the SiV centers and the phonon modes. The central phonon frequency $\omega_0 = \Delta_j + \delta_j$ is assumed to be fixed by compensating small inhomogeneities in the $\Delta_j$ by the detunings $\delta_j = \omega^j_d - (\Delta_j - \omega^j_B)$.

By adiabatically eliminating the fast decaying amplitudes $b_j$, 
we derive effective equations of motion for the slowly varying amplitudes $c_i(t)$. 
From this derivation, detailed in~\cite{SI}, we obtain for each qubit amplitude
\begin{align}\label{Eq:EOM}
	\dot c_j(t) = 
	-&\frac{\gamma_j(t)}{2} c_j(t) - \sum_n\sqrt{\frac{\gamma_{j,n}(t)}{2}}e^{-i\theta_j(t)}\Phi^{\rm in}_{j,n}(t),
\end{align}
where $\gamma_j(t) = \sum_n \gamma_{j,n}(t)$ is the effective decay rate of state $|2\rangle_j$ and
\begin{align}
	\gamma_{j,n}(t)  = \frac{\Omega_j^2(t)}{4\delta_j^2 + \Gamma_j^2(\omega_0)} \Gamma_{j,n}(\omega_0).
	\label{Eq:GammaSpin}
\end{align}
 Assuming $0\leq \Omega(t)/2\pi < 70$ MHz and $\delta/2\pi= 100$ MHz, this rate can be tuned between $\gamma_j=0$ and a maximal value of $\gamma_{\rm max}/2\pi\approx 250$ kHz, which is still fast compared to the expected bare dephasing times $T_2^*=10- 100\,\mu$s  of the qubit state ~\cite{Sukachev2017}.  At the same time, the large detuning $\delta \gg \Gamma(\Delta)$ ensures that any residual  scattering of phonons from an undriven defect is strongly suppressed~\cite{SI}.

\begin{figure}[t]
	\begin{center}
	\includegraphics[width=\columnwidth]{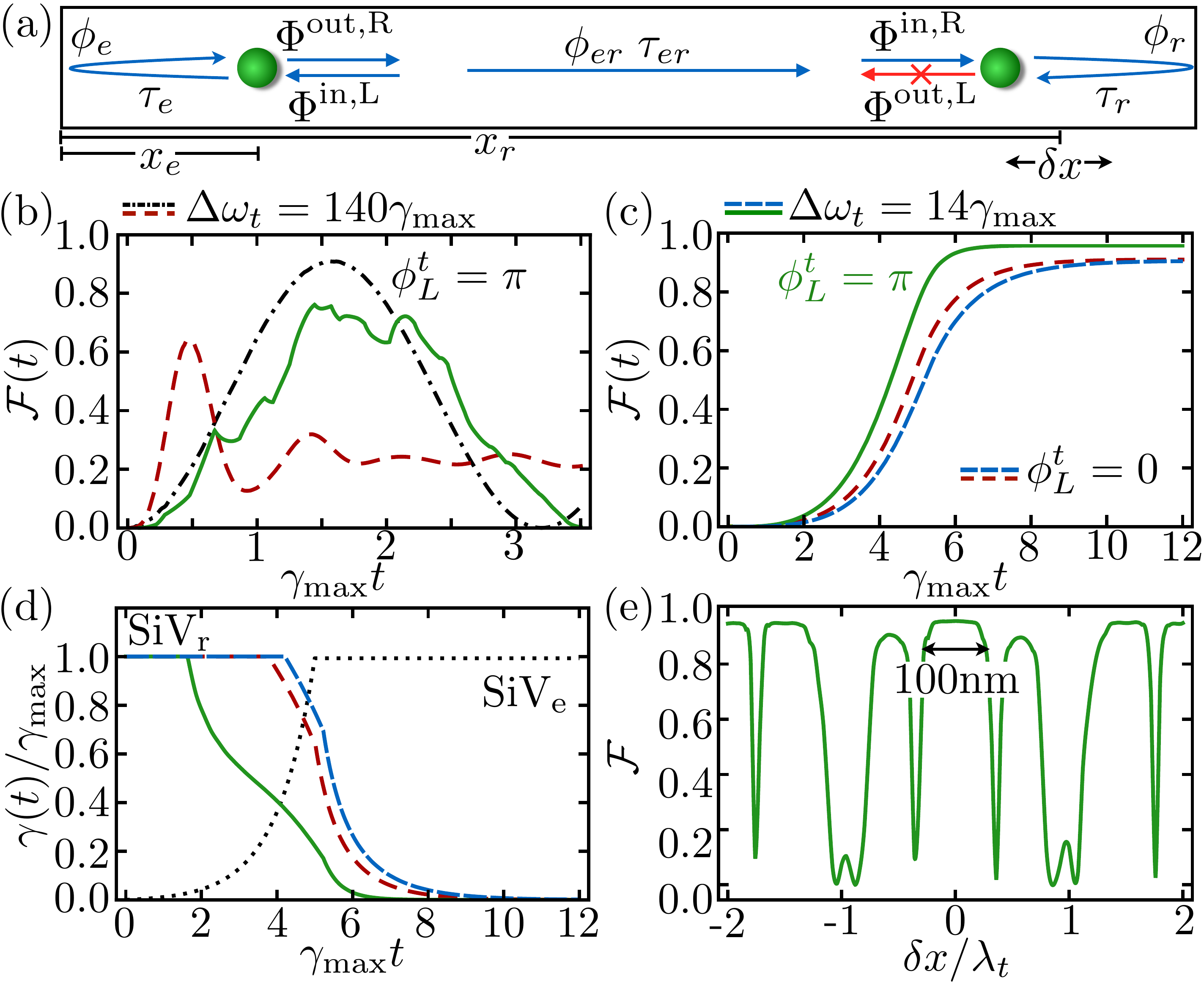}
	\end{center}
	\vspace{-0.5cm} 
	\caption{State transfer protocol. 
	(a) Schematics showing the relevant fields, retardation times and propagation phases. 
	(b) State transfer fidelity for constant rates $\gamma_e(t) = \gamma_r(t) = \gamma_{\rm max}$.
	The case of a single resonant mode (red dashed line; $\phi^t_L = 0$, $\phi^l_L = \pi$) is compared to the off-resonant case (dot-dashed black line; $\phi^t_L = \phi^l_L = \pi$) for $L \sim 100\,\mu$m ($\Delta\omega_t/\gamma_{\rm max} = 140$).
	The full green line represents the long-waveguide counterpart of the off-resonant scenario, where $L \sim 1$ mm ($\Delta\omega/\gamma_{\rm max} = 14$). 
	(c)-(d) Protocol using slowly-varying control pulses ($t_p\gamma_{\rm max} = 1$) where $\Phi^{\rm out, L}_{r,t}(t)$ is completely suppressed. The dashed blue line corresponds to the long waveguide counter part of the dashed red line. 
	For (b)-(d), the two defects are equally coupled to both modes, $\phi^{n}_{e} = \phi^{n}_{r} = \pi$ and $\beta^n_{e} = \beta^n_{r} = 0.5$.
	(e) Plot of the state transfer fidelity for varying positions of the receiving SiV center. For this plot $\phi^t_L = \phi^l_L = \pi$ and a maximal transfer time of $12\gamma_{\rm max}^{-1}$ have been assumed.
	 In all plots, we considered defects near the boundaries where $\tau_e = \tau_r \approx 0$. To illustrate the effect of phonon losses  a boundary reflectivity of $R = 0.92$ has been assumed, which corresponds to $Q \approx 5 \times 10^4$ in the cavity limit.}
\label{Fig:StateTransfer}
\end{figure}

The last term in Eq.~\eqref{Eq:EOM}, where $\Phi^{\rm in}_{j,n}=\Phi^{\rm in,L}_{j,n}+\Phi^{\rm in,R}_{j,n}$, describes the coupling of an SiV center to the left- (L) and right- (R) incoming fields $\Phi^{\rm in, R/L}_{j,n}$, which themselves are related to the corresponding outgoing fields by~\cite{QuantumNoise}
\begin{align}
	\Phi^{\rm out, R/L}_{j,n}(t) =   \Phi^{\rm in, R/L}_{j,n}(t) + \sqrt{\frac{\gamma_{j,n}(t)}{2}}c_j(t)e^{i\theta_j(t)}.
	\label{Eq:InOutRel}
\end{align} 
Together with Eq.~\eqref{Eq:EOM}, these input-output relations specify the local dynamics at each node and must be supplemented by a set of propagation relations for all fields [cf. Fig.~\ref{Fig:StateTransfer}(a)]. As an example, for $x_r>x_e$, the right propagating fields obey 
$\Phi^{\rm in, R}_{r, n}(t) = \Phi^{\rm out, R}_{e, n}(t - \tau^n_{er})e^{i\phi^n_{er}}$, where $\tau^n_{er}=(x_r-x_e)/v_n$ and $\phi^n_{er} = k_n(x_r-x_e)$ are the respective propagation times and  phases. 
Reflections at the boundaries lead to a retarded interaction of each center with its own emitted field. For example,  $\Phi^{\rm in, R}_{e,n}(t) =  -\sqrt{R_n}\Phi^{\rm out, L}_{e, n}(t - \tau^n_e)e^{i\phi^n_e}$, where $\tau^n_e = 2x_e/v_n$ and $\phi^n_e = 2k_n x_e$, and the reflectivity $R_n\leq 1$ has been introduced to model losses. The combined set of time-nonlocal equations for the SiV amplitudes can be solved numerically 
for given positions $x_j$ and pulses $\gamma_{j,n}(t)$. Since any deterministic phase acquired during the protocol can be undone by a local qubit rotation, we identify $\mathcal{F}(t)=|c_r(t)|^2$ with the fidelity of the transfer, which  exceeds the classical bound for $\mathcal{F}>2/3$~\cite{Massar1995}.

\emph{Quantum state transfer.}---In Fig.~\ref{Fig:StateTransfer}(b) we first consider constant rates $\gamma_{j,n}(t) = \gamma_{\rm max}/2$, in which case a state transfer is achieved over multiple round-trips of the emitted wave-packet.  For $L\sim 100\,\mu$m, the round-trip times $2L/v_n$ are still short compared to $\gamma_{\rm max}^{-1}$ and we recover the standing-wave picture with splittings $\Delta\omega_n = \pi v_n / L$ between consecutive $k$-modes. When only the transverse mode is resonant, [i.e., $\phi^{t}_L = \phi^{t}_e + \phi^{t}_r + 2\phi^{t}_{er} = 2\pi n$, while $\phi^{l}_L = (2m+1)\pi$] and for maximal coupling [$\phi^{l}_e = \phi^{l}_r = (2m+1)\pi$], we observe damped oscillations with a fast frequency $\tilde g= \sqrt{\gamma_{\rm max}\Delta\omega_t/2\pi} \approx 2\pi \times 1.2$ MHz and decay rate $\kappa= -\frac{\Delta\omega}{\pi}\log R \approx 2\pi\times0.93$ MHz. This result is expected from a single-mode description of the waveguide~\cite{SI}, and is recovered here as a limiting case of our general framework. The losses from multiple imperfect reflections at the boundaries can be partially suppressed at the expense of a slightly slower transfer by detuning the SiV centers from the closest mode by $\delta_0> \tilde g$. In this case the SiV centers communicate via an exchange of virtual phonons and $\kappa \rightarrow \kappa (\tilde g/\delta_0)^2$. For a maximal detuning $\delta_0 = \Delta \omega_t/2$, the transfer fidelity scales approximately as $\mathcal{F}\simeq R- \pi^2/(8T_2^*\gamma_{\rm max})$~\cite{SI}. For $T_2^*\approx 100\mu$s and $R>0.99$, which can be achieved, for example, by phononic Bragg mirrors~\cite{Maldovan2013}, gate fidelities of $\mathcal{F}\gtrsim0.99$ are possible.

As illustrated by the solid line in Fig.~\ref{Fig:StateTransfer}(b), the simple cavity picture fails for longer waveguides, where multi-mode and propagation effects become non-negligible. In Fig.~\ref{Fig:StateTransfer}(c) we illustrate a more general and more robust protocol, where the phonons ideally travel the waveguide only once. Here, the emission is gradually turned on with a fixed pulse $\gamma_e(t)/\gamma_{\rm max}={\rm min}\{1,e^{(t-5t_p)/t_p}\}$, while $\gamma_r(t)$ and $\theta_r(t)$ are constructed numerically by minimizing at every time step the back-reflected transverse field $|\Phi^{\rm out, L}_{r,t}|$. 
For slow pulses, $\gamma_{\rm max}t_p \gg1$, a perfect destructive interference between the field reflected from the boundary and the field emitted by the receiving center can be achieved, i.e., $\Phi^{\rm in,L}_{r,t}(t) + \sqrt{\gamma_{r,t}(t)/2}c_r(t)e^{i\theta_r(t)}=0$.
For a single branch ($\beta_t = 1$) this results in a complete suppression of the signal traveling back to the emitting center so that for $R=1$ and negligible retardation effects, a perfect state transfer can be implemented~\cite{Habraken2012, Cirac1997,Jahne2007,Korotkov2011}.
Fig.~\ref{Fig:StateTransfer}(c) shows that this approach also leads to high transfer fidelities under more general conditions, where all propagation effects are taken into account and multiple independent channels participate in the transfer.
Importantly, since there are no resonances building up, this strategy is independent of $L$ and can be applied for short and long waveguides equally well.

In the examples shown in Fig.~\ref{Fig:StateTransfer}(b)-(d), the SiV centers are placed at positions near the ends of the waveguide, where the effective emission rate $\tilde\gamma_{j,n}(t) = 2\gamma_{j,n}(t)\sin^2(\phi^n_{j}/2)$~\cite{Dorner2002} into both modes is maximal.
Fig.~\ref{Fig:StateTransfer}(e) shows the achievable transfer fidelities when the position of the receiving center is varied over several wavelengths. We observe plateaus of high fidelity extending over $\sim 100$ nm, interrupted by a few sharp dips arising from a complete destructive interference, i.e.~$\phi_{r}\approx \pi$. This position insensitivity, even in a multi-channel scenario, can be understood from a more detailed inspection of the outgoing fields $\Phi^{\rm out, L}_{r,l}$~\cite{SI} and makes the transfer protocol consistent with uncertainties of $\delta x <50$ nm achieved with state-of-the-art implantation techniques~\cite{Schroder2016}.
 

\begin{figure}[t]
	\begin{center}
	\includegraphics[width= 0.9\columnwidth]{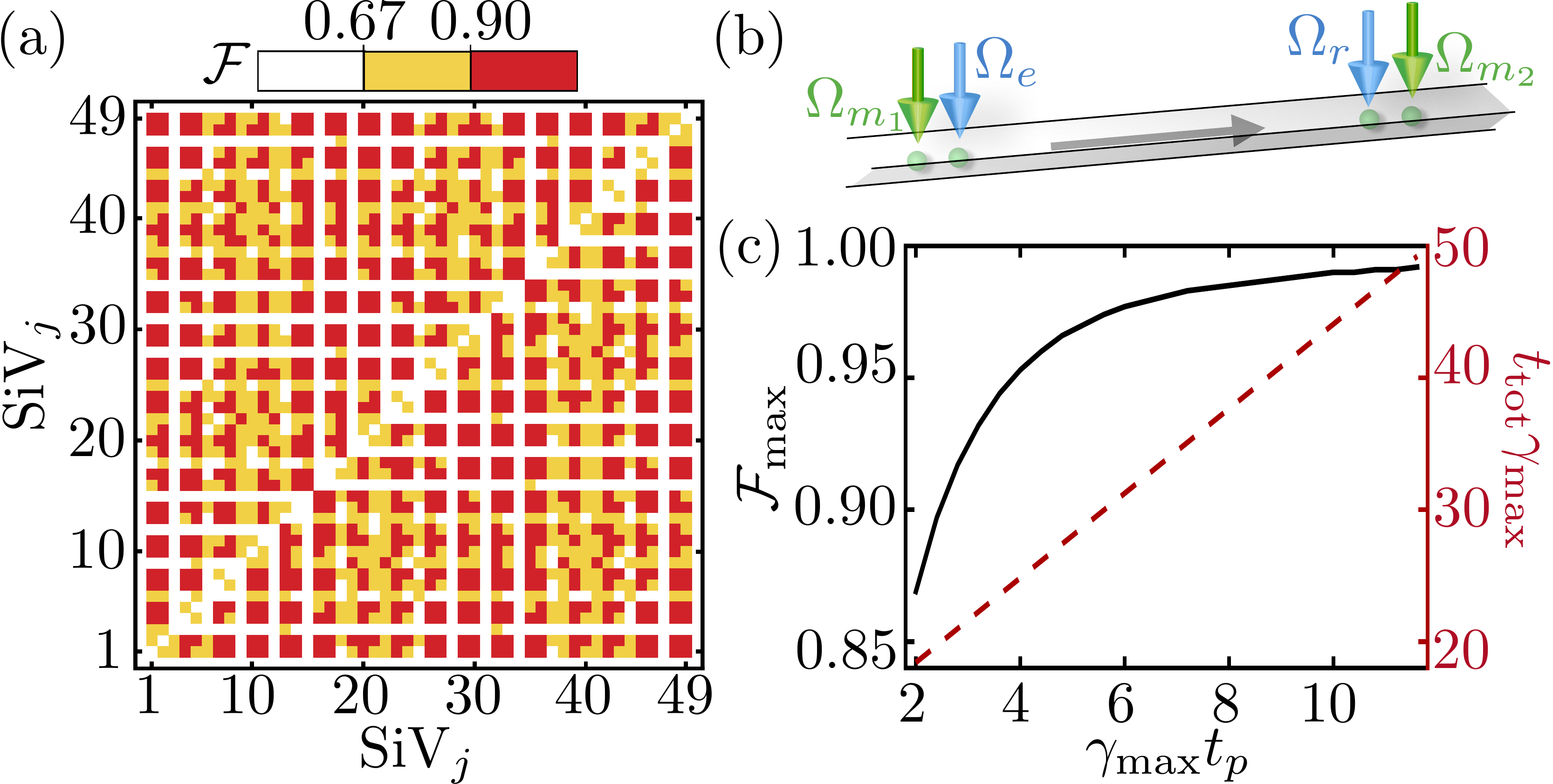}
	\end{center}
	\vspace{-0.5cm} 
	\caption {Scalability. (a) Quantum connectivity matrix for 49 SiV centers equally spaced in a $500\,\mu$m long waveguide. 
	The maximal protocol time is fixed to $12\gamma_{\rm max}^{-1}$ and $R = 0.92$.
	(b) State-transfer protocol in an infinite waveguide where the outermost  SiV centers act as switchable mirrors. The full black line shows the fidelity as a function of the characteristic time of the emitting pulse, $t_p$, while the dashed red curve shows the total protocol time, $t_{\rm tot}$.
For the considered positions the propagation phases between each center and their neighboring mirror defect is $\pi$ for both phonon branches and $\Delta\phi_{er} = 2\pi\times n$.
In all simulations, a constant decay rate for the mirror defects, $\gamma_{m_1}(t) = \gamma_{m_2}(t) = \gamma_{\rm max}$, maximizes the fidelity.}
\label{Fig:Scalability}
\end{figure}


\textit{Scalability.}---In Fig.~\ref{Fig:Scalability} (a), we consider a waveguide of length $L=500\,\mu$m containing $49$ SiV centers  spaced by 
$\Delta x=10\,\mu$m  to allow individual addressing by optical or microwave fields. The resulting quantum connectivity matrix, i.e.~the achievable state transfer fidelity between each pair, shows that apart from a few exceptions, most centers can be connected efficiently and that in principle, the operation of large scale networks is possible. By using phononic bandstructure engineering~\cite{SafaviNaeini2011,Maldovan2013}, single mode~\cite{Patel2017} or chiral phononic waveguides~\cite{Brendel2017}, the transfer fidelities can be further increased beyond the basic scenario considered here. In practice, propagation losses and elastic phonon scattering will set additional limitations for the overall size of the network. In Fig.~\ref{Fig:Scalability} (b), we show a general strategy to overcome these limitations by separating the whole waveguide into smaller segments using additional `mirror centers'. Here the two outermost SiV centers simply reflect the incoming phonon wavepacket~\cite{Shen2005}, and thus create an effective cavity within the waveguide~\cite{Zhou2008,Chang2012}. This is illustrated in Fig.~\ref{Fig:Scalability} (b), where we plot the resulting state transfer fidelity for two SiV centers localized inside this effective cavity.  
For transfer pulses that are long compared to $\gamma_{\rm max}^{-1}$, the outmost centers act as almost perfect mirrors, such that even in an infinite waveguide state transfer protocols within reconfigurable sections of the network can be implemented.

\textit{Conclusion.}---We have shown how an efficient coupling between individual SiV centers and propagating phonons in a diamond waveguide can be realized and used for quantum networking applications. By employing direct spin-phonon couplings in the presence of a transverse magnetic field~\cite{SrujanPrep} or defect-phonon interactions in other materials~\cite{Soykal2011,Rusko2013,Ramos2013}, many of the described techniques could also be adapted for lower phonon frequencies $\sim5-10$ GHz, where many advanced phononic engineering methods are already available. When combined with local  control operations involving adjacent nuclear spins as quantum memories~\cite{Waldherr2014,Rogers2014, Childress2007,Reiserer2016}, the set of all these techniques provides a realistic approach for a scalable quantum information processing platform with spins in solids.

\textit{Acknowledgements.}---This work was supported by the Austrian Science Fund (FWF) through the SFB FoQuS, Grant No. F40, the START Grant No. Y 591-N16, ONR MURI on Quantum Optomechanics (Award No. N00014-15-1-2761), STC Center for Integrated Quantum Materials (NSF Grant No. DMR-1231319), NSF EFRI ACQUIRE (Award No. 5710004174), CUA, NSF and Vannevar Bush Fellowship.

\end{document}


\title{Supplemental material: Phonon networks with SiV centers in diamond waveguides}
\author{M.-A. Lemonde$^1$, S. Meesala$^2$, A. Sipahigil$^3$, M. J. A. Schuetz$^4$, M. D. Lukin$^4$, M. Loncar$^2$, P. Rabl$^1$}
\affiliation{$^1$ Vienna Center for Quantum Science and Technology,
Atominstitut, TU Wien, 1040 Vienna, Austria}
\affiliation{$^2$ John A. Paulson School of Engineering and Applied Sciences, Harvard University, 29 Oxford Street, Cambridge, MA 02138, USA}
\affiliation{$^3$  Institute for Quantum Information and Matter and Thomas J. Watson, Sr., Laboratory of Applied Physics, California Institute of Technology, Pasadena, California 91125, USA}
\affiliation{$^4$  Department of Physics, Harvard University, Cambridge, Massachusetts 02138, USA}

\date{\today}

\maketitle

\section{Electronic structure of the \SiV~center} \label{SI:EStructure}

As described in the main text, the electronic ground state of the negatively charged SiV consists of a single unpaired hole with spin $S=1/2$, which can occupy one of the two degenerate orbital states $|e_x\rangle$ or $|e_y\rangle$. Within the ground state subspace and in the presence of a static  external magnetic field $\vec B$, the energy structure is determined by a spin-orbit interaction, a Jahn-Teller (JT) effect and the Zeeman splittings. The resulting Hamiltonian reads ($\hbar=1$)~\cite{Hepp2014, HeppThesis}
\begin{align} \label{Eq:HSiV}
H_\textrm{SiV}  = -\lambda_\textrm{SO}L_z S_z + H_\textrm{JT}  + f \gamma_L B_z L_z +\gamma_s \vec B \cdot \vec{S}.
\end{align}
Here, $L_z$ and $S_z$ are the projections of the dimensionless angular momentum and spin operators $\vec{L}$ and $\vec{S}$ onto 
the symmetry axis of the center, which we assume to be aligned along the $z$-axis. $\lambda_\textrm{SO} > 0$ is the spin-orbit coupling while $\gamma_L$ and $\gamma_s$ are the orbital and spin gyromagnetic ratio. respectively. The parameter $f\approx 0.1$ accounts for the reduced orbital Zeeman effect in the crystal lattice. Note that within the ground-state subspace spanned by $|e_x\rangle$ and $|e_y\rangle$, only $L_z$ is non-zero.
Within this basis and for an external magnetic field $\vec B = B_0 \vec z$, the different contributions of Eq.~\eqref{Eq:HSiV} read
\begin{align}
	(\omega_B-\lambda_{\rm SO}\hat{L}_z)\hat{S}_z  + \hat{H}_{\rm JT} =
	\frac{1}{2}\begin{bmatrix}
	\omega_B & i\lambda_{\rm SO} \\
	-i\lambda_{\rm SO} & \omega_B
	\end{bmatrix} 
	\otimes
	\begin{bmatrix}
	1 & 0 \\
	0 & -1
	\end{bmatrix}
	+ 
	 \begin{bmatrix}
	\Upsilon_x & \Upsilon_y  \\
	\Upsilon_y & -\Upsilon_x
	\end{bmatrix} \!\otimes\! 
	\begin{bmatrix}
	1 & 0 \\
	0 & 1
	\end{bmatrix}.
\label{Eq:HSiVMatrix}
\end{align}
Here, $\Upsilon_x$ ($\Upsilon_y$) denotes the strength of the Jahn-Teller coupling along $x$ ($y$) and $\omega_B = \gamma_sB_0$ is the Zeeman energy. From this point, we neglect for simplicity the effect of the reduced orbital Zeeman interaction ($\sim f\gamma_L B_0$), which does not affect any of the results in the main text.  Diagonalizing Eq.~\eqref{Eq:HSiVMatrix} leads to the eigenstates 
\begin{align}
\begin{split}
        \vert 1 \rangle & = \left(\cos\theta \vert e_x \rangle - i  \sin\theta e^{-i\phi} \vert e_y \rangle\right)\vert\!\downarrow \rangle, \\
	\vert 2 \rangle & = \left(\cos\theta \vert e_x \rangle + i\sin\theta e^{i\phi}\vert e_y \rangle\right)\vert\!\uparrow \rangle, \\
	\vert 3 \rangle & = \left(\sin\theta \vert e_x \rangle + i  \cos\theta e^{-i\phi}\vert e_y \rangle\right)\vert\!\downarrow \rangle, \\
	\vert 4 \rangle & = \left(\sin\theta \vert e_x \rangle - i  \cos\theta e^{i\phi}\vert e_y \rangle\right)\vert\!\uparrow \rangle,
\end{split}
\end{align} 
where
\begin{align}
	\tan(\theta) = \frac{2\Upsilon_x + \Delta}{\sqrt{\lambda_{\rm SO}^2 + 4\Upsilon_y^2}}, \qquad
	\tan(\phi) = \frac{2\Upsilon_y}{\lambda_{\rm SO}}.
\end{align}
The corresponding eigenenergies are
\begin{align}
	E_{3,1} = (-\omega_B \pm \Delta)/2, \quad E_{4,2} = (\omega_B \pm \Delta)/2,
\end{align}
with $\Delta=\sqrt{\lambda_{\rm SO}^2+4(\Upsilon_x^2 + \Upsilon_y^2)}\approx 2\pi \times 46$ GHz.
Since $\Upsilon_{x,y} \ll \lambda_{\rm SO}$ (cf.~Ref.~\cite{Hepp2014}), we can neglect the small distortions of the orbital states by the JT effect and therefore use the approximation $|1\rangle\approx |e_-,\downarrow\rangle$, $|2\rangle\approx |e_+,\uparrow\rangle$, $|3\rangle\approx |e_+,\downarrow\rangle$ and $|4\rangle\approx |e_-,\uparrow\rangle$, which corresponds to $\theta = \pi/4$ and $\phi = 0$.

\section{Phonon waveguide}

In the main text we consider a diamond phonon waveguide with a cross section $A$ and a length $L \gg \sqrt{A}$. Within the frequency range of interest, the phonon modes can be modelled as elastic waves with a displacement field $\vec u(\vec r,t)$ obeying the equation of motion for a linear, isotropic medium~\cite{ClelandBook},
\begin{equation}\label{eq:EOM_Displacement}
\rho \frac{\partial^2}{\partial t^2} \vec u =  (\lambda+\mu) \vec \nabla ( \vec \nabla\cdot \vec u) + \mu \vec \nabla^2 \vec u,
\end{equation}
or in terms of the individual components
\begin{equation}
\rho \frac{\partial^2}{\partial t^2} u_k =  (\lambda+\mu)\sum_m  \frac{\partial^2 u_m}{\partial x_k \partial x_m} + \mu \sum_m \frac{\partial^2 u_k}{\partial x_m^2}.
\end{equation}
Here, $\rho$ is the mass density and the Lam\'e constants 
\begin{equation}
\lambda =\frac{\nu E}{(1+\nu)(1-2\nu)},\qquad \mu = \frac{E}{2(1+\nu)},
\end{equation}
can be expressed in terms of the Young's modulus $E$ and  the Poisson ratio $ \nu$. In our calculations and finite element method (FEM) simulations, we use $\rho=3500$ kg/m$^3$, $E=1050$ GPa and $\nu=0.2$.

By assuming periodic boundary conditions, the equations of motion can be solved by the general ansatz
\begin{equation}\label{eq:Ansatz_u}
\vec u(\vec r,t)= \frac{1}{\sqrt{2}}\sum_{k,n}  \vec u^\perp_{n,k}(y,z) \left[ A_{n,k}(t)    e^{ikx} + A^*_{n,k}(t) e^{-ikx}  \right], 
\end{equation}
where $k=2\pi/L\times m$ is the wavevector along the waveguide direction $x$, and the index $n$ labels the different phonon branches. The amplitudes $A_{n,k}(t)$ are oscillating functions obeying $\ddot A_{n,k}(t)+\omega_{n,k}^2A_{n,k}(t)=0$, and the mode frequencies $\omega_{n,k}$ and the transverse mode profile $\vec u^\perp_{n,k}(x,y)$ are in general obtained from a numerical solution of the Eq.~\eqref{eq:EOM_Displacement}.
The $\vec u^\perp_{n,k}(x,y)$ are orthogonal and normalized to  
\begin{equation}
\frac{1}{A} \int dy dz   \, \vec u^\perp_{n,k}\cdot\vec u^\perp_{\beta,k} =\delta_{n,\beta}.
\end{equation}

\subsection{Quantization of the displacement field}

Eq.~\eqref{eq:EOM_Displacement} can be derived from the Lagrangian 
\begin{equation}
L= \int d^3 r  \,\left[  \frac{\rho}{2}\dot{\vec u}^2  -  \frac{(\lambda+\mu)}{2} \sum_{k,m} \frac{\partial u_k}{\partial x_m} \frac{\partial u_m}{\partial x_k} - \frac{\mu}{2} \sum_{k,m} \left(\frac{\partial u_k}{\partial x_m}\right)^2  \right]. 
\end{equation}
After inserting the eigenmode decomposition in Eq.~\eqref{eq:Ansatz_u}, the Lagrangian reduces to a set of harmonic modes:
\begin{equation}
L(\{Q_{n,k}\},\{\dot Q_{n,k}\})= \sum_{k,n}  \frac{M}{2} \dot Q_{n,k} \dot Q_{n,-k} -  \frac{1}{2} M \omega_{n,k}^2 Q_{n,k}Q_{n,-k},
\end{equation}
where $M=\rho AL$ and $Q_{n,k}=(A_{n,k}+A_{n,-k}^*)/\sqrt{2}$. From this simplified form, we readily obtain the canonical momenta $P_{n,k}=\partial L/\partial \dot Q_{n,k}=M \dot Q_{n,-k}$, and the Hamiltonian operator  
\begin{equation}
H_{\rm ph}= \sum_{k,n}  \frac{P_{n,k} P_{n,-k}}{2M}  +  \frac{1}{2} M \omega_{n,k}^2 Q_{n,k}Q_{n,-k},
\end{equation}
where $Q_{n,k}$ and $P_{n,k}$ are now operators obeying the canonical commutation relations, $[Q_{n,k},P_{n,k}]=i\hbar \delta_{n,n'}\delta_{kk'}$. Finally, we write 
\begin{equation}
Q_{n,k}= \sqrt{\frac{\hbar}{2M\omega_{n,k}}}\left(a^\dag_{n,k}  + a_{n,-k}\right),
\qquad P_{n,k}= i \sqrt{\frac{\hbar M\omega_{n,k}}{2}}\left(a^\dag_{n,k}  -a_{n,-k}\right), 
\end{equation}
in terms of annihilation and creation operators. We obtain 
\begin{equation}
H_{\rm ph}= \sum_{k,n}  \hbar \omega_{n,k} a^\dag_{n,k} a_{n,k},
\end{equation}
and the quantized displacement field
\begin{equation}
\vec u(\vec r)= \sum_{k,n} \sqrt{\frac{\hbar}{2M\omega_{n,k}}} \vec u^\perp_{n,k}(y,z) \left(   a_{n,k}  e^{ikx} + a^\dag_{n,k} e^{-ikx}  \right). 
\label{Eq:Disp}
\end{equation}

\section{Coupling to phonon modes}

Strain coupling arises from the change in Coulomb energy of the electronic states due to displacement of the atoms forming the defect. For small displacements and in the Born-Oppenheimer approximation, the energy shift is linear in the local distortion and can be written as
\begin{equation}
	H_{\rm strain} = \sum_{ij} V_{ij} \epsilon_{ij}.
\end{equation}
Here, $V$ is an operator acting on the electronic states of the SiV defect and $\epsilon$ is the strain tensor defined as
\begin{equation}
	\epsilon_{ij} = \frac{1}{2}\left( \frac{\partial u_i}{\partial x_j} + \frac{\partial u_j}{\partial x_i}\right),
\end{equation}
with $u_{1}$ ($u_{2}$, $u_{3}$) representing the quantized displacement  field  along $x_1 = x$ ($x_2 = y, x_3 = z$) at the position of the SiV center [cf.~Eq.~\eqref{Eq:Disp}].
The axes are defined as in Fig.~1 of the main text, i.e.~the symmetry axis of the defect is along $z$ while the waveguide is along $x$.

The exact form of the strain interaction Hamiltonian in the basis of the electronic states of the SiV defect is obtained by projecting the strain tensor on the irreducible representations of the $D_{3d}$ group, i.e.
\begin{equation}
	H_{\rm strain} = \sum_{r} V_r \epsilon_r, \label{H.str}
\end{equation}
where $r$ denotes the irreducible representations. One can show that the only contributing representations are the one-dimensional representation $A_{1g}$ and the two-dimensional representation $E_{g}$~\cite{HeppThesis}. As a consequence, strain can couple independently to orbitals within the $E_g$ and $E_u$ manifolds, but these manifolds cannot be mixed. Focusing only on the ground state, the terms in Eq.~\eqref{H.str} are \cite{Meesala2017}
\begin{eqnarray}
\epsilon_{A_{1g}} &=& t_{\perp}(\epsilon_{xx}+\epsilon_{yy}) + t_{\parallel}\epsilon_{zz} \nonumber\\
\epsilon_{E_{gx}} &=& d(\epsilon_{xx}-\epsilon_{yy}) + f\epsilon_{zx} \\
\epsilon_{E_{gy}} &=& -2d\epsilon_{xy} + f\epsilon_{yz}\nonumber
\end{eqnarray}
Here, $t_{\perp}, t_{\parallel}, d, f$ are the strain-susceptibilities with $f/d \sim 10^{-4}$, and the subscript $g$ is used to denote the ground state manifold. The effects of these strain components on the electronic states are described by
\begin{eqnarray}
	V_{A_{1g}} &=& \ket{e_x}\bra{e_x} + \ket{e_y}\bra{e_y}\nonumber\\
	V_{E_{gx}} &=& \ket{e_x}\bra{e_x} - \ket{e_y}\bra{e_y}\\
	V_{E_{gy}} &=& \ket{e_x}\bra{e_y} + \ket{e_y}\bra{e_x}\nonumber
\end{eqnarray}

Since coupling to symmetric local distortions ($\sim \epsilon_{A_{1g}}$) shifts  all ground states equally, it has no relevant effects in this work and can thus be dropped. 
Finally, if we write the strain Hamiltonian using the basis spanned by the eigenstates of the spin-orbit coupling $\{|e_{-} \rangle, |e_{+} \rangle \}$, we find
\begin{align}
H_{\rm strain}  =  \epsilon_{E_{gx}} \left(L_{-}+L_{+} \right)  - i\epsilon_{E_{gy}} \left(L_{-}-L_{+} \right), \nonumber
\end{align}
where $L_+ = L_-^\dag = \ket{3}\bra{1} + \ket{2}\bra{4}$ is the orbital raising operator within the ground state. Further, we notice that the transitions $L_+, L_-$ have circularly polarized selection rules in the $E_g$ strain components.

We now assume that the SiV high symmetry axis $z$ is oriented orthogonal to the phonon propagation direction $x$ (practically, this can be realized with [110]-oriented diamond waveguides). By decomposing the local displacement field as in Eq.~\eqref{Eq:Disp}, and after making a rotating wave approximation, the resulting strain coupling can be written as  
\begin{equation}
	H_\textrm{strain} \simeq \frac{1}{\sqrt L}\sum_{n,k}  \left[ (g_{n,k} J^\uparrow_+ + g^*_{n,-k} J^\downarrow_+) a_{n,k} e^{ikx} + {\rm H.c.}\right],
	\label{Eq:Hstrain}
\end{equation}
where $J^\uparrow_+ = \ket{3}\bra{1}$, $J^\downarrow_+ = \ket{4}\bra{2}$ and
\begin{align}
	g_{n, k} & = d\sqrt{\frac{\hbar k^2}{2\rho A \omega_{n,k}}}\frac{1}{|k|}\left[ \left(ik u^{\perp,x}_{n,k} + ik\frac{f}{d}\frac{u^{\perp,z}_{n,k}}{2} + \frac{f}{d}\frac{\partial_z u^{\perp,x}_{n,k}}{2} - \partial_y u^{\perp,y}_{n,k}\right) - i\left(ik u^{\perp,y}_{n,k} + \partial_y u^{\perp,x}_{n,k} +\frac{f}{d}\frac{\partial_y u^{\perp,z}_{n,k}}{2} + \frac{\partial_z u^{\perp,y}_{n,k}}{2}\right) \right], \nonumber \\
	& \equiv d\sqrt{\frac{\hbar k^2}{2\rho A \omega_{n,k}}}\xi_{n,k}(y,z).
	\label{Eq:gstrain}
\end{align} 
Here, $u^{\perp,i}_{n,k}$ represents the $i$-th component of the displacement pattern $\vec u^{\perp}_{n,k}(y,z)$. The first four terms in the square bracket correspond to $E_{g_x}$ deformations, while the last four correspond to $E_{g_y}$ deformations. We note from Eq.~\eqref{Eq:Hstrain} that due to circularly polarized selection rules, it is possible to have different coupling rates to left or right propagating phonons and that this directionality is reversed, when the spin character of the states involved in the phononic transition is flipped. This is due to the particular energy-state ordering in which $E_{\downarrow, +} > E_{\downarrow, -}$ while $E_{\uparrow, +} < E_{\uparrow, -}$. However, the waveguide phonon modes considered in this work are approximately linearly polarized with predominantly $E_{gx}$ strain, and hence have identical coupling rates for both propagation directions (and spin projections). Therefore, the strain Hamiltonian reduces to
\begin{equation}
	H_\textrm{strain} \simeq \frac{1}{\sqrt L}\sum_{n,k}  g_{n,k} J_+ a_{n,k} e^{ikx} + {\rm H.c.}
	\label{Eq:HstrainFinal}
\end{equation}

\section{Spin-phonon interface}

In this section, we present in more details two different driving schemes for transferring spin-states encoded in the SiV ground-state to propagating phonons. 
We first consider the scenario depicted in the main text that utilizes a microwave drive within the ground-state subspace. Furthermore, we present a second approach via optical Raman transitions to the excited states, which can be a useful alternative to microwave magnetic fields. 
For simplicity, we first focus on a single SiV center in an infinite waveguide.

\subsection{Microwave driving fields}

The starting point is the Hamiltonian of a single driven SiV center coupled via strain to the phonon modes of the diamond waveguide, 
\begin{equation}
	H = H_{\rm SiV} + H_{\rm ph} + H_{\rm drive} + H_{\rm strain}.
\end{equation}
By moving into the interaction picture with respect to 
\begin{equation}
	H_0 = \sum_{n,k} \omega_0 a_{n,k}^\dag a_{n,k} +  \omega_B \ket{2}\bra{2} + \omega_0 \ket{3}\bra{3} + (\omega_0 + \omega_B)\ket{4}\bra{4},
	\label{Eq:RotFrame}
\end{equation}
we obtain the new Hamiltonian $\tilde{H} = e^{i H_0t } H e^{-i H_0t} - H_0$ given by
\begin{align}
	\tilde{H} = &\sum_{n,k} (\omega_{n,k} - \omega_0)a_{n,k}^\dag a_{n,k} - \delta( \ket{3}\bra{3} + \ket{4}\bra{4}) \nonumber \\
	& + \left[ \frac{\Omega(t)e^{i\theta(t)}}{2}(\ket{3}\bra{2} + e^{2i\omega_Bt}\ket{4}\bra{1})
	+ \frac{1}{\sqrt{L}}\sum_{n,k}g_{n,k}e^{ikx}a_{n,k}(\ket{3}\bra{1} + \ket{4}\bra{2}) + \rm{H.c.} \right].
	\label{Eq:HFull}
\end{align}
Here, $\omega_0 = \Delta + \delta$ is the central frequency of the emitted phonon wavepackets and  $\Omega(t)$ and $\theta(t)$ are the strength and phase of the external driving field,  respectively.

In this rotating frame, the ansatz for the single-excitation wavefunction reads
\begin{equation}
	|\psi(t)\rangle = \alpha\ket{1,0} + \beta\big[ c(t) |2\rangle\langle 1|   + b(t) |3\rangle\langle 1| + \sum_{n,k}  c_{n,k}(t)a^\dag_{n,k}\big]  \ket{1,0},
\end{equation}
where $\ket{1,0}$ is the ground state with the SiV center in state $|1\rangle$ and no phonons in the waveguide. 
Note  that this ansatz does not capture the off-resonant transition $\ket1\rightarrow \ket4$ produced by the drive [i.e., the term $\sim e^{2i\omega_Bt}$ in Eq.~\eqref{Eq:HFull}].
We estimate its effect below and show that for the parameters considered in this work, it can be neglected.

From the Schr{\"o}dinger equation $\partial_t |\psi(t)\rangle = -i\tilde H(t) |\psi(t)\rangle$, we obtain the equations of motion for the amplitudes,
\begin{align}
\begin{split}
	\dot c_{n, k}(t) & = -i(\omega_{n, k} - \omega_0)c_{n, k}(t) - i \frac{1}{\sqrt L}g^*_{n, k}e^{-ikx}b(t), \\
	\dot c(t) & = -i\frac{\Omega(t)e^{-i\theta(t)}}{2}b(t), \\
	\dot b(t) & = i\delta b(t) - i\frac{\Omega(t) e^{i\theta(t)}}{2}c(t) - i\frac{1}{\sqrt{L}}\sum_{n, k} g_{n, k}e^{ikx}c_{n, k}(t).
\end{split}
\label{Eq:EOMSingleSiV}
\end{align}
The solution for the propagating phonons reads
\begin{align}
	c_{n, k}(t) = e^{-i(\omega_{n,k} - \omega_0)(t - t_0)}c_{n,k}(t_0) - \frac{i}{\sqrt L}g^*_{n, k}e^{-ikx}\int_{t_0}^t d\tau e^{-i(\omega_{n,k} - \omega_0)(t-\tau)}b(\tau),
\end{align}
where $t_0$ is an arbitrary time before any phonons interacted with the SiV center.
Plugging this result back into the equation for the excited state amplitude, we obtain
\begin{align}
	\dot b(t) = i\delta b(t) - i\frac{\Omega(t) e^{i\theta(t)}}{2}c(t) - i\frac{1}{\sqrt{L}}\sum_{n, k} g_{n, k}e^{ikx}e^{-i(\omega_{n,k} - \omega_0)t}c_{n,k}(t_0)
	  - \frac{1}{L}\sum_{n, k} |g_{n, k}|^2\int_{t_0}^t d\tau e^{-i(\omega_{n,k} - \omega_0)(t-\tau)}b(\tau).
\end{align}
In the present case where the SiV center is driven by phonons of frequencies close to $\omega = 0$ ($\omega_0$ in the lab frame), also the amplitude $b(t)$ is slowly varying, allowing us to perform a standard Markov approximation~\cite{QuantumNoise}. This results in
\begin{equation}
	\dot b(t) = \left[i\delta - \frac{\Gamma(\omega_0)}{2}\right] b(t) - i\frac{\Omega(t) e^{i\theta(t)}}{2}c(t) 
	- \sum_n \sqrt{\frac{\Gamma_{n}(\omega_0)}{2}}\big[\Phi^{\rm in, L}_{n, \omega_0}(t) + \Phi^{\rm in, R}_{n, \omega_0}(t)\big],
\end{equation}
with the phonon-induced decay rate
\begin{equation}
	\Gamma(\omega) = \frac{2\pi}{L}\sum_{n, k} |g_{n, k}|^2\delta(\omega - \omega_{n, k}) = 2\frac{|g_{n}|^2}{v_n},
	\label{Eq:GammaPH}
\end{equation}
and the input field
\begin{align}
	\Phi^{\rm in, L/R}_{n, \omega}(t) = i\sum_{k>0} \sqrt{\frac{v_n}{L}} e^{\mp i k x}e^{-i(\omega_{n, k} - \omega)(t - t_0)}c_{n, k}(t_0).
	\label{Eq:InputField}
\end{align}
Here, the group velocity $v_n = d\omega_{n,k}/dk$ and the coupling constant $g_{n} = g_{n, k}$ are evaluated at $\omega_0$ and are considered constant over the frequency range of interest [$\delta\omega \sim \Gamma(\omega)$] around $\omega_0$. 

\begin{figure}
\centering
\includegraphics[width= 0.8 \columnwidth]{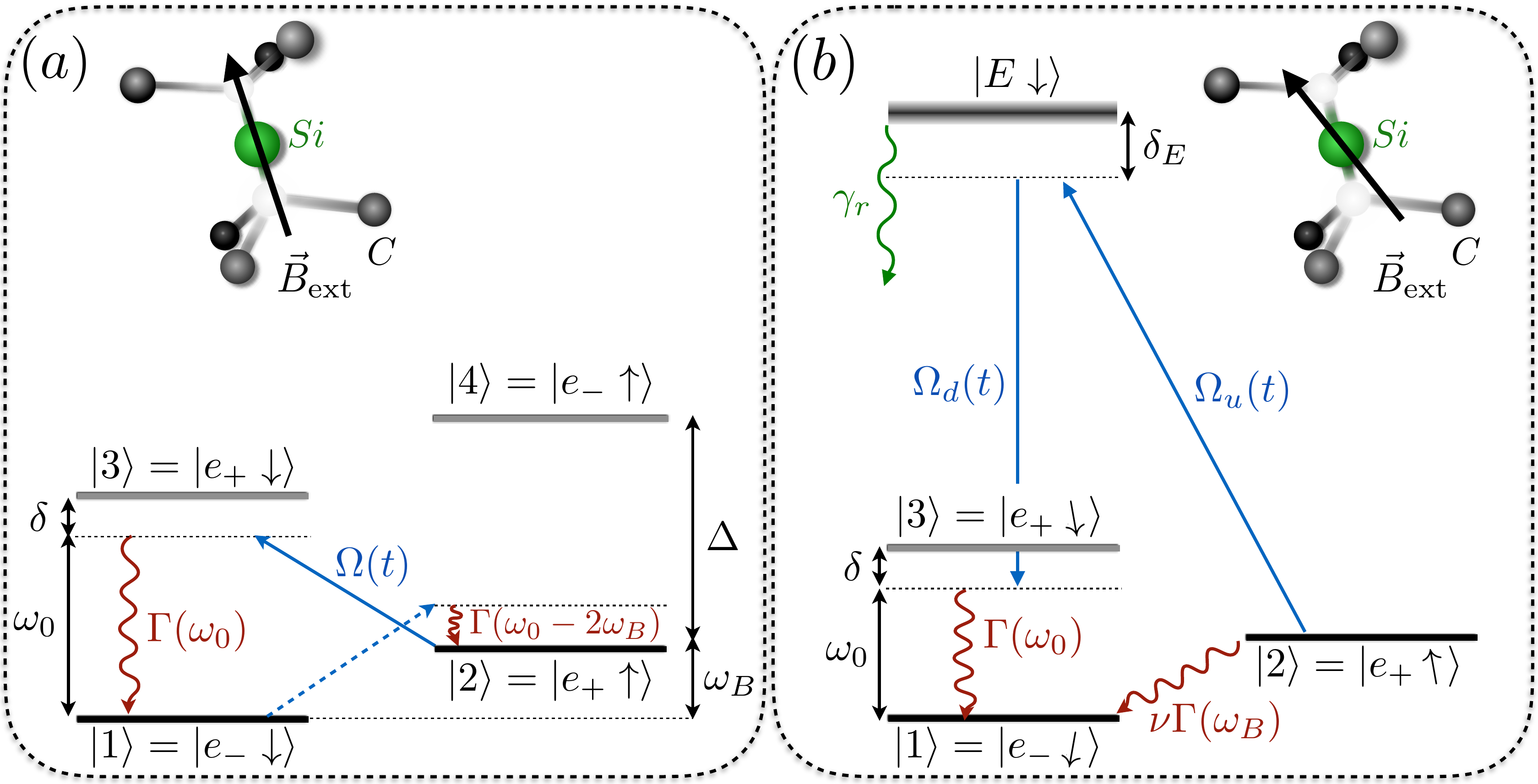}
\caption{Comparison of the two driving schemes to implement the spin-phonon interface. (a) A microwave magnetic field drives the transition $\ket2 \rightarrow \ket3$, and state $\ket3$ subsequently decays to state $\ket1$ by emitting a propagating phonon at frequency $\omega_0$. The process that drives the transition $\ket1 \rightarrow \ket4$ is strongly off-resonant for large Zeeman energy $\omega_B \gg \delta$ and can be neglected.
(b) The transition $\ket2 \rightarrow \ket3$ is now driven by two optical fields via the excited state $\ket {e_u \downarrow}$ of the SiV center. In that case, the external magnetic field has to be tilted from the symmetry axis of the defect. As a consequence, it opens a decoherence channel via the direct transition $\ket2 \rightarrow \ket1$.}
\label{Fig:OptDrive}
\end{figure} 

We now make further simplifications and consider weak and slowly-varying driving fields, i.e., $\Omega(t) \ll |\delta + i\Gamma(\omega_0)/2|$, $\dot\Omega(t)/|\delta + i\Gamma(\omega_0)/2|$ and $ \dot\theta(t) \ll |\delta + i\Gamma(\omega_0)/2|$.
Given those constraints, one can adiabatically eliminate the higher-energy state, i.e.~$\dot b(t) = 0$, and obtain
\begin{equation}
	\dot c(t) = -\left[ i\omega_s(t) + \frac{\gamma(t)}{2} \right]c(t) 
	-\sum_n\sqrt{\frac{\gamma_n(t)}{2}}e^{-i\bar\theta(t)}\big[\Phi_n^{\rm in, L}(t) + \Phi_n^{\rm in, R}(t)\big],
\end{equation}
with
\begin{align}
	\omega_s(t) = \frac{\Omega^2(t)}{4}\frac{\delta}{\delta^2 + \Gamma^2(\omega_0)/4},
	\quad
	\gamma(t) = \sum_n \gamma_n(t)= \frac{\Omega^2(t)/4}{\delta^2 + \Gamma^2(\omega_0)/4}\sum_n\Gamma_{n}(\omega_0),
	\quad
	\bar\theta(t) = \theta(t) + \arctan\left[\frac{\Gamma(\omega_0)}{2\delta}\right]
	\label{Eq:Gamma}
\end{align}
The AC-Stark-shift $\omega_s(t)$ can be compensated by a corresponding (slow) adjustment of the driving frequency $\omega_{d}(t)$ to keep $\omega_0$ constant during the entire driving protocol. Doing so and omitting the constant shift of the drive phase $\bar\theta(t) \rightarrow \theta(t)$ for simplicity, one recovers the form introduced in Eq.~(7) of the main text.

\subsubsection{Residual driving of the transition $\ket1 \rightarrow \ket4$}

As described by Eq.\eqref{Eq:HFull}, the microwave field also drives the transition $|1\rangle \rightarrow |4\rangle$. One can estimate the rate at which this process takes place by applying the same procedure as above starting from the following  ansatz
\begin{equation}
	|\psi(t)\rangle = \alpha\ket{2,0} + \beta\big[ c(t) |1\rangle\langle 2|   + b(t) |4\rangle\langle 2| + \sum_{n,k}  c_{n,k}(t)a^\dag_{n,k}\big]  \ket{2,0},
\end{equation}
where $\ket{2,0}$ represents the SiV center in state $|2\rangle$ and no phonons in the waveguide. 
Doing so, one finds
\begin{equation}
\dot c(t) =  -\left[ i\tilde\omega_s(t) + \frac{\tilde\gamma(t)}{2}\right]c(t) 
	- \sum_n\sqrt{\frac{\tilde\gamma_{n}(t)}{2}}e^{-i\theta(t)}[\Phi^{\rm in, L}_{n, \omega_0-2\omega_B}(t) + \Phi^{\rm in, R}_{n, \omega_0-2\omega_B}(t)], 
\end{equation}
with the input fields defined in Eq.~\eqref{Eq:InputField},
the AC-Stark-shift and the effective transfer rate 
\begin{align}
	\tilde\omega_s(t) = \frac{\Omega^2(t)}{4}\frac{\delta - 2\omega_B}{(\delta - 2\omega_B)^2 + \Gamma^2(\omega_0 - 2\omega_B)/4},
	\qquad
	\tilde\gamma(t) = \frac{\Omega^2(t)/4}{(\delta - 2\omega_B)^2 + \Gamma^2(\omega_0 - 2\omega_B)/4}\sum_n\Gamma_{n}(\omega_0 - 2\omega_B).
\end{align}
As a consequence, the drive also allows the SiV states to flip from $\ket 1$ to $\ket 2$ by emitting a phonon at frequency $\omega_0 - 2\omega_B$. For large Zeeman splittings, the rate of this process
\begin{equation}
	\frac{\tilde\Gamma(t)}{\Gamma(t)} \sim \frac{\delta^2}{(\delta - 2\omega_B)^2}\frac{\Gamma_{\rm ph}(\omega_0 - 2\omega_B)}{\Gamma_{\rm ph}(\omega_0)}
\end{equation}
 is strongly suppressed, as long as $\Gamma_{\rm ph}(\omega_0 - 2\omega_B) \simeq \Gamma_{\rm ph}(\omega_0)$. Therefore, care must be taken to avoid band edges at frequencies near $\omega_0 - 2\omega_B$.

\subsection{Optical Raman driving schemes}

We now present an alternative driving scheme that makes use of the electronically excited states via an optical  two-tone Raman transition, as depicted in Fig.~\ref{Fig:OptDrive} (b). The corresponding Hamiltonian reads
\begin{equation}
	H = H_{\rm SiV} + H_{\rm ph} + H_{\rm drive} + H_{\rm strain} + H_{\rm rad},
	\label{Eq:HtotOpt}
\end{equation}
where $H_{\rm rad}$ captures the radiative decay of the excited states, $H_{\rm drive}$ describes the optical driving of the excited state and $H_{\rm SiV}$ now includes a component of the magnetic field perpendicular to the symmetry axis of the defect (e.g.~along $x$). The perpendicular field allows one to couple opposite-spin states via optical driving fields \cite{Meesala2017,Rogers2014,Pingault2014}.

\subsubsection{Effects of a weak perpendicular magnetic field}

Focusing only on the ground-state subspace and the only relevant excited state $\ket{E \downarrow}$, 
\begin{align}
	H_{\rm SiV} & = -\lambda_{\rm SO}L_z S_z + \omega_B S_z + \omega_x B_x S_x, \nonumber \\
	 & = -\frac{\Delta + \omega_B}{2}\ket{e_-\downarrow}\bra{e_-\downarrow} - \frac{\Delta - \omega_B}{2}\ket{e_+\uparrow}\bra{e_+\uparrow} \label{Eq:Hopt} \\
	 & \qquad + \frac{\Delta - \omega_B}{2}\ket{e_+\downarrow}\bra{e_+\downarrow} + \frac{\Delta + \omega_B}{2}\ket{e_-\uparrow}\bra{e_-\uparrow} + \omega_E\ket{E \downarrow}\bra{E \downarrow},\nonumber 
\end{align}
where $\omega_B = \gamma_s B_z$, $\omega_x \equiv \gamma_s B_x$, and $\omega_E$ is the energy of the excited state.
As described in the first section, Eq.~\eqref{Eq:Hopt} neglects the orbital Zeeman effect and the distortion of the orbital states due to the JT effect.

In the limit of weak perpendicular magnetic field $\omega_x / |\Delta - \omega_B| \ll 1$, a small mixing between opposite-spin states occurs, leading to new eigenstates:
\begin{gather}
	H_{\rm SiV} = \sum_{i=1,4}\omega_i \ket i\bra i + \omega_E\ket{E \downarrow}\bra{E \downarrow},
	\qquad \Rightarrow 
	\left\lbrace \begin{matrix}
		\,\,\,\,\ket 1 \approx \ket{e_- \downarrow} - \eta_+ \ket{e_- \uparrow}, 
		\qquad \omega_1 \approx  -\frac{\Delta + \omega_B}{2} - \frac{\eta_+\omega_x}{2} \\
	\,\,\,\,\ket 2 \approx \ket{e_+ \uparrow} - \eta_-\ket{e_+ \downarrow},
	\qquad \omega_2 \approx  -\frac{\Delta - \omega_B}{2} - \frac{\eta_-\omega_x}{2} \\
	\ket 3 \approx \ket{e_+ \downarrow} +\eta_-\ket{e_+ \uparrow}, 
		\qquad \omega_3 \approx  \frac{\Delta - \omega_B}{2} + \frac{\eta_-\omega_x}{2} \\
	\ket 4 \approx \ket{e_- \uparrow} + \eta_+\ket{e_- \downarrow},
	\qquad \omega_4 \approx  \frac{\Delta + \omega_B}{2} + \frac{\eta_+\omega_x}{2}
	\end{matrix}\right.,
\end{gather} 
with 
\begin{equation}
	\eta_\pm \equiv \frac{1}{2} \frac{\omega_x}{\Delta \pm \omega_B}, 
	\qquad
	\eta = \eta_- + \eta_+.
\end{equation}
Note that due to the larger spin-orbit interaction in the excited state ($\sim 250$GHz) the effect of $B_x$ on $\ket{E \downarrow}$ can be neglected.
In this new basis, the strain interaction given in Eq.~\eqref{Eq:HstrainFinal} becomes
\begin{align}
	H_\textrm{strain} = \sum_{n,k}  g_{n, k}a_{n,k}\left[ J_+  +\eta(\ket4\bra3 - \ket2\bra1)\right] + \rm{H.c.}
\end{align}
As a consequence, a magnetic field which is not perfectly aligned with the symmetry axis of the SiV center induces a finite strain coupling between states $\ket1 \leftrightarrow \ket2$ and $\ket3 \leftrightarrow \ket4$. 

Within the same basis, the optical driving fields with frequencies $\omega_u$ and $\omega_d$ are described by
\begin{align}
	H_{\rm drive} & = \left( \frac{\Omega_d(t)e^{i\theta_d(t)}}{2}e^{-i\omega_d t} + \frac{\Omega_u(t)e^{i\theta_u(t)}}{2}e^{-i\omega_u t}\right)\ket{E \downarrow} \bra{e_+ \downarrow} + {\rm H.c.} \nonumber \\
	& = \frac{\Omega_d(t)e^{i\theta_d(t)}}{2}e^{-i\omega_d t}\ket{E \downarrow}\bra 3 - \frac{\Omega_u(t)e^{i\theta_u(t)}}{2}\eta_-e^{-i\omega_u t}\ket{E \downarrow}\bra 2 + {\rm H.c.}
\end{align}
The last line is obtained by making a rotating wave approximation valid for large frequency mismatch between the two drives, i.e.~$|\omega_d - \omega_u| \gg \Omega_{d,u}$.

\subsubsection{Effective 3-level system}

To extract the effective rate at which the spin state is transferred to propagating phonons and estimate the dephasing rates due to the radiative decay of the excited state and the direct strain coupling between state $\ket1$ and $\ket2$, we apply the  same procedure as the previous section. This time, we work in the rotating frame with respect to 
\begin{equation}
	H_0 = \sum_{n,k} \omega_0a_{n,k}^\dag a_{n,k} +  \sum_i \omega_i\ket{i}\bra{i} + \delta( \ket{3}\bra{3} +  \ket{4}\bra{4}) + \delta_E\ket{E\downarrow}\bra{E\downarrow},
\end{equation}
and with the drives detuned such that $\omega_d = \omega_E - \omega_3 + \delta_E - \delta$ and $\omega_u = \omega_E - \omega_2 + \delta_E$ [see Fig.~\ref{Fig:OptDrive} (b)].
From the low-excitation ansatz
\begin{equation}
	|\psi(t)\rangle = \alpha\ket{1,0} + \beta\big[ c(t) |2\rangle\langle 1|  + b(t) |3\rangle\langle 1| + E(t) |E\downarrow\rangle\langle 1| + \sum_{n,k}  c_{n,k}(t)a^\dag_{n,k}\big]  \ket{1,0},
\end{equation}
we derive the Schr\"odinger equations for the time-dependent coefficients. We approximate the effects of the dipole interaction ($H_{\rm rad}$) by including a finite lifetime of the excited state $\ket{E \downarrow}$ in the form of a radiative decay $\Gamma_{\rm rad}$ ($\sim 100$ MHz), i.e.
\begin{equation}
	\dot E(t) = \left( i\delta_E - \frac{\Gamma_{\rm rad}}{2}\right) E(t) - \frac{i}{2}\big[\Omega_d(t)e^{i\theta_d(t)}b(t) - \eta_-\Omega_u(t)e^{i\theta_u(t)}c(t)\big].
\end{equation}

We focus on the limit of weak optical drives $\Omega_{u,d}(t) \ll |\delta_E + i\Gamma_{\rm rad}/2|, |\delta + i\Gamma/2|$ so that we can adiabatically eliminate the excited state [$\dot E(t) = 0$]. Within the Markov approximation, this leads to an effective 3-level system, where
\begin{align}
\begin{split}
	\dot b(t) & = -\left\lbrace i\big[ \omega_d(t) - \delta \big] + \frac{\gamma_{\rm rad}^d(t)}{2} + \frac{\Gamma(\omega_0)}{2} \right\rbrace b(t)
	+ i \frac{\Omega_{\rm eff}(t)}{2}e^{i[ \theta_{\rm eff}(t) - \phi_{\rm NH}]}c(t)
	+\sqrt{\frac{\Gamma_n(\omega_0)}{2}}\big[ \Phi^{\rm in, L}_{n, \omega_0}(t) + \Phi^{\rm in, R}_{n, \omega_0}(t) \big], \\
	\dot c(t) & = -\left\lbrace i\omega_u(t) + \frac{\gamma_{\rm rad}^u(t)}{2} + \frac{\eta^2\Gamma(\omega_B)}{2} \right\rbrace c(t)
	+ i \frac{\Omega_{\rm eff}(t)}{2}e^{-i[ \theta_{\rm eff}(t) + \phi_{\rm NH}]}b(t)
	+\sqrt{\frac{\eta^2\Gamma_n(\omega_B)}{2}}\big[ \Phi^{\rm in, L}_{n, \omega_B}(t) + \Phi^{\rm in, R}_{n, \omega_B}(t) \big],
\end{split}
	\label{Eq:3Level}
\end{align}
with the phonon-induced decay rate and input fields defined in Eqs.~\eqref{Eq:GammaPH} and \eqref{Eq:InputField} respectively. 
The phase $\phi_{\rm NH} = \arctan(\Gamma_{\rm rad}/2\delta_E)$ comes from the radiative decay and can be neglected for large detunings $\delta_E \gg \Gamma_{\rm rad}$. Note that in Eqs.~\eqref{Eq:3Level}, we have neglected higher-order virtual processes that couple states $\ket2$ and $\ket3$ via strain interaction that are strongly off-resonant for $|g_{n,k}|^2/|\Delta - \omega_B| \ll \Gamma(\omega_0), \Gamma(\omega_B)$. 

At this stage, we recover the 3-level system utilized within the magnetic driving scheme described above, except for the AC-Stark shifts of states $\ket2$ and $\ket3$,
\begin{equation}
	\omega_u(t) = \frac{\eta_-^2}{4}\frac{\Omega_u^2(t)}{\delta_E^2 + \Gamma_{\rm rad}^2/4}\delta_E, 
	\qquad
	\omega_d(t) = \frac{1}{4}\frac{\Omega_d^2(t)}{\delta_E^2 + \Gamma_{\rm rad}^2/4}\delta_E,
\end{equation}
respectively, and additional decay channels. One of the new loss mechanism comes from the radiative decay of the excited state $\ket E$, which affects both states $\ket2$ and $\ket3$ with respective rates
\begin{equation}
	\gamma^u_{\rm rad}(t) = \frac{\eta_-^2}{4}\frac{\Omega_u^2(t)}{\delta_E^2 + \Gamma_{\rm rad}^2/4}\Gamma_{\rm rad},
	\qquad
	\gamma^d_{\rm rad}(t) = \frac{1}{4}\frac{\Omega_d^2(t)}{\delta_E^2 + \Gamma_{\rm rad}^2/4}\Gamma_{\rm rad},
\end{equation}
while the finite strain coupling between states $\ket1$ and $\ket2$ also induces an addition decay channel with rate $\eta^2\Gamma({\omega_B})$ and incoming noise $\Phi^{\rm in}_{\omega_B}(t)$.
Finally, the effective Rabi frequency driving the transition $\ket2 \rightarrow \ket 3$ is given by
\begin{equation}
	\Omega_{\rm eff}(t) = \frac{\eta_-}{2}\frac{\Omega_d(t)\Omega_u(t)}{\sqrt{\delta_E^2 + \Gamma_{\rm rad}^2/4}},
	\qquad
	\theta_{\rm eff}(t) = \theta_u(t) - \theta_d(t).
\end{equation}

The viability of this scheme resides in the relative importance of the loss mechanisms compared to the coherent dynamics. More precisely, the phonon-assisted transfer rate of state $\ket3$ has to overcome its radiative decay, i.e.~$\Gamma(\omega_0)  \gg \gamma^d_{\rm rad}(t)$, while the other loss mechanisms as to be overcome by the final spin-state transfer rate, i.e.~$\gamma(t) \sim \frac{\Omega_{\rm eff}^2(t)}{\delta^2}\Gamma(\omega_0) \gg \gamma^u_{\rm rad}(t), \eta^2\Gamma(\omega_B)$. As an example (all rates are divided by $2\pi$), for Rabi frequencies $\Omega_d \approx \Omega_u/2 \sim 2.5$ GHz, detunings $\delta_E \sim 30$ GHz and $\delta \sim 30$ MHz, a radiative decay rate $\Gamma_{\rm rad} \sim 100$ MHz, and a ratio $\eta_- \sim 0.1$, the different maximal rates are 
\begin{equation}
	\gamma^d_{\rm rad} \sim 150\, {\rm kHz} \ll \Gamma(\omega_0) \sim 2\, {\rm MHz}, \qquad
	\gamma^u_{\rm rad} \sim 7\, {\rm kHz},\,
	\eta^2\Gamma(\omega_B) \sim 40\, {\rm kHz} \ll
	\gamma \sim 250\, {\rm kHz}.
\end{equation}
Here, we use $\Gamma(\omega_B) \sim \Gamma(\omega_0) \sim 1$ MHz. Using optical driving should thus be a viable route to achieve a fully controllable spin transfer into a propagating phonon.

\section{Input-output formalism}

In this section, we extend the previous calculations to multiple SiV defects and recover the input-output relations stated in the main text. We first start by considering an infinite waveguide, where we explicitly derive how the input field of a given center is related to the output field of the others. In this scenario, we estimate the effects of phonon scattering by undriven defects. Finally, we close the section by considering the effects of waveguide boundaries.

\subsection{Infinite waveguides}

The coherent dynamics of the SiV ensemble is governed by the Hamiltonian given in Eq.~(2) in the main text, which in the rotating frame defined in Eq.~\eqref{Eq:RotFrame} reads
\begin{align}
	H = \sum_{n,k} (\omega_{n,k} - \omega_0)a_{n_k}^\dag a_{n_k} 
	+ \sum_{j} H^{(j)}_{\rm SiV}
	+ \frac{1}{\sqrt{L}}\sum_{j,n,k}\left[g^j_{n,k}e^{ikx_j}a_{n,k}J^j_+ + \rm{H.c.} \right],
\end{align}
with 
\begin{equation}
	H^{(j)}_{\rm SiV} = -\delta_j( \ket{3}\bra{3} + \ket{4}\bra{4}) + \left[ \frac{\Omega_j(t)e^{i\theta_j(t)}}{2}\ket{3}\bra{2} + {\rm H.c.} \right].
\end{equation}
In this frame, the single-excitation ansatz considered in the main text, $|\psi(t)\rangle =[\alpha \mathds{1} +\beta C^\dag(t)] \ket{\bar 1,0}$, is now defined with
\begin{equation}
C^\dag(t) = \sum_{j=e,r}  \big[ c_j(t) |2\rangle_j\langle 1|   + b_j(t) |3\rangle_j\langle 1| + \sum_{n,k}  c_{n,k}(t)a^\dag_{n,k}\big],
\label{Eq:LEAnsatz}
\end{equation}
where $\ket{\bar 1,0}$ is the ground state with all SiV centers in state $|1\rangle$ and no phonon in the waveguide. In Eq.~\eqref{Eq:LEAnsatz}, we only kept the two driven centers, i.e.~the emitting (e) and receiving (r) one.

The equations of motion for the different amplitudes are
\begin{align}
\begin{split}
	\dot c_{n, k}(t) & = -i(\omega_{n, k} - \omega_0)c_{n, k}(t) - i \frac{1}{\sqrt L}(g^e_{n, k})^*e^{-ikx_e}b_{e}(t) - i \frac{1}{\sqrt L}(g^r_{n, k})^*e^{-ikx_r}b_{r}(t), \\
	\dot c_{j}(t) & = -i\frac{\Omega_j(t)e^{-i\theta_j(t)}}{2}b_{j}(t), \\
	\dot b_{j}(t) & = i\delta_j b_{j}(t) - i\frac{\Omega_j(t) e^{i\theta_j(t)}}{2}c_{j}(t) - i\frac{1}{\sqrt{L}}\sum_{n, k} g^j_{n, k}e^{ikx_j}c_{n, k}(t).
\end{split}
\end{align}
We again apply the same procedure as in the previous sections, i.e., we first exactly solve the equation for $c_{n,k}(t)$, insert the solution in the equation for $b_j(t)$ and then perform a Markov approximation. Doing so for the receiving defect and taking $x_r > x_e$, we obtain
\begin{align}
	\dot b_{r}(t) = &\left[i\delta_r - \frac{\Gamma_r(\omega_0)}{2}\right] b_{r}(t) - i\frac{\Omega_r(t) e^{i\theta_r(t)}}{2}c_{r}(t)
	 - i\sum_n\sum_{k} \sqrt{\frac{\Gamma_{r,n}(\omega_0)}{2}}\sqrt{\frac{v_n}{L}} e^{i k x_r}e^{-i(\omega_{n, k} - \omega_0)(t-t_0)}c_{n, k}(t_0)  \nonumber \\
	&-\sum_{n} \sqrt{\frac{\Gamma_{r,n}(\omega_0)}{2}\frac{\Gamma_{e,n}(\omega_0)}{2}}e^{ik_n(x_r-x_e)} b_{e}(t - \tau^n_{er}),
\end{align}
with $\tau_{er}^n = (x_r - x_e)/v_n$. 
We recall that $g_{n,k}$ is taken to be real without loss of generality and $t_0$ is a time in the past before the two SiV defects have interacted with incoming wavepackets.

The final step is to adiabatically eliminate the higher-energy state [$\dot b_{j}(t) = 0$] for both SiV centers and insert the result in the equation for $c_{j}(t)$, which leads to the final form
\begin{equation}
	\dot c_r(t) = -\left[i\omega_{s,r} + \frac{\gamma_r(t)}{2}\right]c_r(t) 
	-\sum_n\sqrt{\frac{\gamma_{r,n}(t)}{2}}e^{-i\bar\theta_r(t)}[\Phi_{r,n}^{\rm in, L}(t) + \Phi_{r,n}^{\rm in, R}(t) + \Phi_{r,n}^{\rm scatt}(t)].
	\label{Eq:EOMSS}
\end{equation}
Here, the AC-Stark shift $\omega_{s,r}$, the effective transfer rate $\gamma_{r,n}(t)$ and the shifted driving phase $\bar\theta_r(t)$ are all defined in Eq.~\eqref{Eq:Gamma} by taking $\Omega \rightarrow \Omega_r$ and $\delta \rightarrow \delta_r$. 
The left- and right-propagating input fields are respectively ($\bar \theta \rightarrow \theta$ for simplicity)
\begin{align}
\begin{split}
	\Phi_{r,n}^{\rm in, L}(t) & = i\sum_{k>0} \sqrt{\frac{v_n}{L}} e^{-i k x_r}e^{-i(\omega_{n, k} - \omega_0)(t - t_0)}c_{n, k}(t_0), \\
	\Phi_{r,n}^{\rm in, R}(t) & = i\sum_{k>0} \sqrt{\frac{v_n}{L}} e^{i k x_r}e^{-i(\omega_{n, k} - \omega_0)(t - t_0)}c_{n, k}(t_0) + \sqrt{\frac{\gamma_{e,n}(t-\tau^n_{er})}{2}}e^{i\theta_e(t-\tau_{er}^n)}e^{ik_n(x_r-x_e)},
\end{split}
\label{Eq:Input}
\end{align}
while $\Phi_{r,n}^{\rm scatt}(t)$ describes back-scattered fields from undriven centers; its expression and effects are described below.
In terms of input-output formalism, we can recast the right-propagating input field as
\begin{align}
	\Phi_{r,n}^{\rm in, R}(t) & = \Phi_{e,n}^{\rm out, R}(t - \tau_{er}^n)e^{i\phi_{er}^n} 
	\qquad\therefore\quad 
	\Phi_{e,n}^{\rm out, R}(t) = \Phi_{e,n}^{\rm in, R}(t) + \sqrt{\frac{\gamma_{e,n}(t)}{2}}e^{i\theta_e(t)},	
\end{align}
with $\phi^n_{er} = k_n(x_r - x_e)$; as stated in the main text.

\subsection{Reflection at the boundaries}

So far, we have considered an infinite waveguide, therefore leading to free propagating wavepackets as input fields $\Phi_{r,n}^{\rm in, L}(t)$ and $\Phi_{e,n}^{\rm in, R}(t)$, as described in Eq.~\eqref{Eq:Input}. For finite waveguides, as in the main text, one needs to specify how the propagating phonons behave at the boundaries. For hard reflections, we have
\begin{equation}
	\Phi_{e,n}^{\rm in, R}(t) = -\sqrt{R_n}\Phi_{e,n}^{\rm out, L}(t-\tau^n_e)e^{i\phi^n_e},
	\qquad
	\Phi_{r,n}^{\rm in, L}(t) = -\sqrt{R_n}\Phi_{r,n}^{\rm out, R}(t-\tau^n_r)e^{i\phi^n_r},
\end{equation}
where the delay times are $\tau^n_e = 2x_e/v_n$, $\tau^n_r = 2(L - x_r)/v_n$ and the phases $\phi^n_e = 2k_n x_e$, $\phi^n_r = 2k_n (L-x_r)$.
We capture losses at those boundaries by introducing the reflectivity $R_n < 1$. By mapping the resulting losses on an exponential decay, one can estimate the corresponding quality factor $Q$, i.e.
\begin{equation}
	R_n = e^{-\kappa_n L/v_n},
	\qquad \Rightarrow \qquad
	Q = \frac{\omega_0}{\kappa_n} = -\frac{\omega_0}{\log(R_n)}\frac{L}{v_n},
	\quad
	\kappa_n = -\log(R_n)\frac{v_n}{L} = -\log(R_n)\frac{\Delta\omega_n}{\pi}.
	\label{Eq:Reflectivity}
\end{equation}
For example, $L = 100\,\mu$m, $v_t = 0.7 \times 10^4$ m/s, $\omega_0 = 2\pi \times 46$ GHz and $R_t = R_l = 0.92$, as in Fig.~3 of the main text, corresponds to $Q \approx 4.95 \times 10^4$.

\subsection{Scattering from undriven centers}

As mentioned above, the last term in Eq.~\eqref{Eq:EOMSS}, 
\begin{equation}
	\Phi^{\rm scatt}_{r,n}(t) = -i\sum_{n'} \frac{1}{\delta_e + i\Gamma_e(\omega_0)/2}\sqrt{\frac{\Gamma_{e,n}(\omega_0)}{2}\frac{\Gamma_{e,n'}(\omega_0)}{2}}\Phi^{\rm out, L}_{r,n}(t - \tau^n_{er} -  \tau^{n'}_{er})e^{i(\phi^n_{er} - \phi^{n'}_{er})},
\end{equation}
represents incoming fields that have been previously emitted by the receiving SiV and scattered back by the emitting center. Note that the amplitude of the scattered field does not depend on the drive applied on the emitting center. Therefore, such scattering process can occur at any defects along the waveguide. 
To avoid unwanted scattering during the state-transfer protocol, it is thus important to always work in the far detuned regime $\delta_j \gg \Gamma_j(\omega_0)$.

\section{State-transfer Fidelity}

In this section, we give additional details about the state-transfer protocol presented in the main text. More precisely, we show how the single-mode limit can be approximately described by a Jaynes-Cumming type interaction and how the fidelity is affected by the difference between the phases gathered by both phonon branches upon propagation in the multimode case.

\subsection{Constant driving of both centers}

\begin{figure}
\centering
\includegraphics[width= 0.8 \columnwidth]{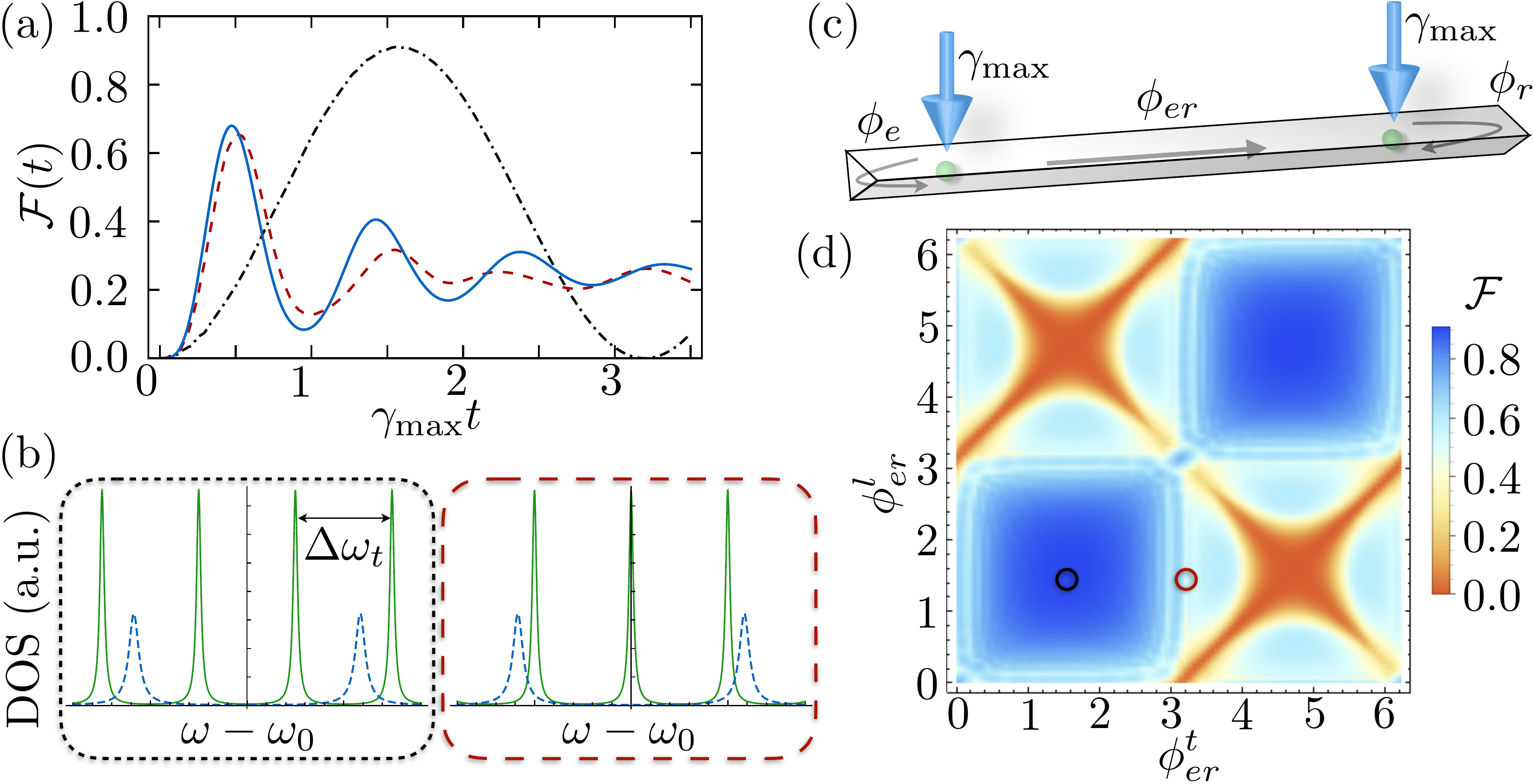}
\caption{State transfer for constant driving of the SiV centers. (a) Fidelity as a function of time for two different scenarios. The black dot-dashed line corresponds to both phonon branches being off-resonant, while the dashed red line corresponds to only the transverse being resonant. In the latter case, the state transfer can be approximate by a single-mode effective model, as shown by the full blue line. (b) Corresponding density of states of the transverse (full green) and longitudinal (dashed blue) mode for the off-resonant and single-mode scenarios. (c) Schematic of the state transfer with constant drives. (d) Fidelity as a function of the phase gathered during propagation between both defects by the transverse ($\phi_{er}^t$) and longitudinal ($\phi_{er}^t$) modes. The red and black circles correspond to the off-resonant and single-mode scenario respectively. For all results, we consider maximal coupling of the two centers to both modes ($\phi_e^n = \phi_r^n = \pi$), a frequency splitting $\Delta\omega_t/\gamma_{\rm max} = 140$ and a reflectivity $R = 0.92$. (see main text)}
\label{Fig:ConsDrive}
\end{figure} 

We first focus on the scenario where the drives on both defects are constant, $\gamma_e(t) = \gamma_r(t) = \gamma_{\rm max}$. In this case, the state transfer is performed over multiple round-trips along the waveguide. For small structures ($L\sim 100\,\mu$m) with high quality factor ($Q\sim10^4$), this results in a state transfer via standing-wave modes that are well-resolved in the frequency domain [cf.~Fig.~\ref{Fig:ConsDrive} (b)].

\subsubsection{Single-mode limit}

For a drive frequency tuned so that $\omega_0$ is near resonant with a single frequency-resolved mode, as shown in the right graph of Fig.~\ref{Fig:ConsDrive} (b), we can neglect the effects of all other modes and use an effective single-mode description. To do so, we redo the quantization procedure outlined above, but using a mode expansion in terms of standing waves. 
We obtain the quantized displacement field 
\begin{equation}
	\vec u(\vec r) = \sum_{k,n}\sqrt{\frac{\hbar}{M\omega_{n,k}}}\vec u^\perp_{n,k}(y,z)\big( a_{n, k} + a^\dag_{n, k} \big)\cos(k x),
\end{equation}
where compared to the plane-wave decomposition [c.f.~Eq.~\eqref{Eq:Disp}], the zero-point fluctuation is increased by a factor $\sqrt2$ and the sum runs over positive $k$ vectors with $\Delta k=\pi/L$. In this standing-wave basis and following a standard rotating wave approximation, the strain coupling reads
\begin{equation}
	H_\textrm{strain} = \sqrt\frac{2}{L}\sum_{n,k}  g_{n,k}\sin(kx) J_+ a_{n,k}  + {\rm H.c.}
	\approx   \sqrt\frac{2}{L}  g_{t,k_t}\sin(k_t x) J_+ a  + {\rm H.c.},
\end{equation}
where $g_{n,k}$ is defined in Eq.~\eqref{Eq:gstrain} as before. The last expression is valid in the single-mode limit where the transverse mode with k-vector $k_t$ is resonant, i.e.~$a = a_{t,k_t}$.

In presence of a far-detuned weak microwave field driving the transition $\ket2 \rightarrow \ket3$ [cf.~Eq.~\eqref{Eq:HFull}], the effective Hamiltonian describing the center as an effective 2-level system coupled to a single phonon mode reads
\begin{equation}
	H = \delta a^\dag a -ig\sigma_+ a + {\rm H.c.}, \qquad {\rm with} \qquad g  = \sqrt{\frac{2}{L}}\frac{g_{t,k_t}\Omega}{2\delta}\sin(k_t x),
\end{equation} 
where $\sigma_+ = \ket2\bra1$ and $\delta = \omega_0 - \omega_{t,k_t}$ is the detuning between the emitted phonon and the standing-wave mode frequency. 
Here, we have adiabatically eliminated the higher energy state $\ket3$ and explicitly considered a time-independent drive.
Generalized to the case of a receiving and an emitting center at positions $x_r$ and $x_e$,  respectively, we obtain
\begin{equation}
	H_{\rm s.m.} =  \delta a^\dag a + g_e\sigma_{e,+} a  + g_r \sigma_{r,+} a + {\rm H.c.}, 
	\qquad \therefore \quad
	g_j  = \sqrt{\frac{2}{L}}\frac{g^j_{t,k_t}\Omega_j}{2\delta_j}\sin(k_t x_j).
	\label{Eq:Hsm}
\end{equation}

In Fig.~\ref{Fig:ConsDrive} (a), we compare time evolution obtained from this effective model with $\delta = 0$ to the full calculation presented in the main text. In the full calculation, the decay rate into the transverse mode, in the limit of large detuning $\delta \gg \Gamma(\omega_0)$, is [cf.~Eq.~\eqref{Eq:Gamma}]
\begin{equation}
\gamma_{j,t} = \frac{\Omega_j^2/4}{\delta_j^2 + \Gamma_j^2(\omega_0)/4}\Gamma_{j,t}(\omega_0) \approx \frac{\Omega_j^2}{2\delta_j^2}\frac{|g^j_{t, k_t}|^2}{v_t}.
\end{equation}
Given that $\gamma_{j,t} = \gamma_{\rm max}/2$, the comparison becomes adequate by using 
\begin{equation}
	|g_j| = \sqrt{\frac{\gamma_{\rm max}\Delta\omega_t}{2\pi}}\sin(k_t x_j) = \sqrt{\frac{v_t\gamma_{\rm max}}{2L}}\sin(k_t x_j),
\end{equation}
as supported by Fig.~\ref{Fig:ConsDrive} (a). 
The discrepancy between the two approaches comes from the contribution of the detuned longitudinal and transverse modes.
 
In order to get more insights regarding the state transfer time and the effect of losses, we now explicitly solve the state evolution under the single-mode dynamics described by Eq.~\eqref{Eq:Hsm}. Focusing on the low-excitation wavefunction
\begin{equation}
	|\psi(t)\rangle = \alpha \mathds{1}\ket0 + \sum_{j=e,r}  \big[ c_j(t)\sigma_{j,+} + c_p(t)a^\dag \big] \ket0,
\end{equation}
the time evolution is given by 
\begin{equation}
	\partial_t c_j(t) = -igc_p(t), \qquad
	\partial_t c_p(t) = -(i\delta - \kappa/2)c_p(t) - ig[c_1(t) + c_2(t)].
\end{equation}
Here, we have considered $g_e = g_r = g$ for simplicity and modeled the loss by a dissipation term for the phononic mode [cf.~eq.~\eqref{Eq:Reflectivity}].
Including the initial conditions $c_e(0) = 1$ and $c_r(0) = c_p(0) = 0$, the solutions read
\begin{align}
\begin{split}
	c_e(t) & = 1 + c\frac{g}{\tilde\omega_-}(e^{-i\tilde\omega_-t} - 1) -  c\frac{g}{\tilde\omega_+}(e^{-i\tilde\omega_+t}-1), \\
	c_r(t) & = -1 + c\frac{g}{\tilde\omega_-}(e^{-i\tilde\omega_-t} +1) -  c\frac{g}{\tilde\omega_+}(e^{-i\tilde\omega_+t} + 1), \\
	c_p(t) & = ce^{-i\tilde\omega_-t} - ce^{-i\tilde\omega_+t}, 
\end{split}
\label{Eq:slns}
\end{align}
with 
\begin{equation}
	\tilde\omega_\pm = \frac{\delta}{2} - i\frac{\kappa}{4} \pm \sqrt{2g^2 + \frac{1}{4}\left(\delta - i\frac{\kappa}{2}\right)^2},
	\qquad {\rm and} \qquad
	c = \frac{1}{2g}\left[ \frac{1}{\tilde\omega_-} - \frac{1}{\tilde\omega_+} \right]^{-1}.
\end{equation}
For the resonant scenario plotted in Fig.~\ref{Fig:ConsDrive} (a) (red dashed and blue full curve), $\delta = 0$ and $\kappa \simeq g$, one gets
\begin{equation}
	|c_r(t)|^2 \approx \frac{1}{4} [1 - \cos(\sqrt{2}gt)e^{-\kappa t/4}]^2.
\end{equation}
The state transfer time is thus $T_g = \frac{\pi}{\sqrt{2}g}$ and the fidelity $F \approx \frac{1}{4}(1 + e^{-\kappa t/4})^2$. For $\Delta\omega_t/\gamma_{\rm max} = 140$ and $R = 0.92$, as in Fig.~\ref{Fig:ConsDrive} (a), it leads to $F\approx 0.68$.

\subsubsection{Multimode limit}

In the limit where both branches are off-resonant, as pictured in the left graph of Fig.~\ref{Fig:ConsDrive} (b), the single-mode picture fails. In that limit, not only the phases $\phi^n_e$ and $\phi^n_r$ that determine the effective coupling strength of the SiV centers to the mode $n$ matter, but also the phase that each mode acquires by traveling the waveguide, $\phi^n_{er}$, becomes relevant. 
In Fig.~\ref{Fig:ConsDrive} (d), we plot the state-transfer fidelity as a function of $\phi^t_{er}$ and $\phi^l_{er}$ for the particular case $\phi^n_e = \phi^n_r = \pi$. We see that the fidelity is maximal when both mode are in-phase, while the fidelity goes to zero when the phase difference is $\Delta\phi_{er} = \pi$.

In the limit case where $\omega_0$ is maximally detuned from the modes of both branches $n$ and that all the phases are identical, we can approximate the dynamics using a strongly detuned single-mode model as described by Eq.~\eqref{Eq:Hsm} with $\delta = \Delta\omega_t/2$ and add independently the contribution of the four closest modes [two per branches, see Fig.~\ref{Fig:ConsDrive} (b)]. 

For $\delta = \Delta\omega_t/2 \gg g$, the solution of Eqs.~\eqref{Eq:slns} becomes
\begin{equation}
	|c_r(t)|^2 = \frac{1}{4}\left\vert1 - e^{(4i\frac{g^2}{\Delta\omega_t} - \frac{g^2}{\delta^2}\kappa)t}\right\vert^2.
\end{equation}
In this single mode case, the state transfer time is $T_g = \frac{\pi\Delta\omega_t}{4g^2}$ and the fidelity reads
\begin{equation}
	\mathcal{F} \approx  1 - \frac{4g^2}{\Delta\omega_t^2}\kappa T_g \approx R.
\end{equation}
This result successfully applies to the case where four modes contribute as in the full calculation shown in Fig.~\ref{Fig:ConsDrive} (a) (black dash-dotted curve). The effects of the four modes are to divide by four the transfer time $T_g/4$, but also leads to four independent dissipative channels, therefore multiplying by four the decay rate.

The total fidelity, including the dephasing rate ($1/T_2^*$) of the SiV centers in the multimode case finally reads
\begin{equation}
	\mathcal{F} \approx R -  \frac{\pi\Delta\omega_t}{16g^2T_2^*}.
\end{equation}

\subsection{Time dependent driving}

In this final section, we focus on protocols where  the drive on the emitting center is gradually turned on with a fixed pulse $\gamma_e(t)/\gamma_{\rm max}={\rm min}\{1,e^{(t-5t_p)/t_p}\}$, while $\gamma_r(t)$ and $\theta_r(t)$ are constructed numerically by minimizing at every time steps the magnitude of the back-reflected transverse field $|\Phi^{\rm out, L}_{t}|$. 

In a scenario where only the transverse branch contributes and where any retardation effects are negligible, this protocol leads to a perfect unidirectional state transfer where all signal emitted toward the receiving center is absorbed.
However, in the more realistic scenario where also the longitudinal field is excited, such a driving scheme does not assure the suppression of the total reflected signal as $|\Phi^{\rm out, L}_{l}|$ can be finite. 
In what follows, we estimate the conditions in which this protocol leads to high-fidelity state transfers in the general multimode case.

In the simplest limit where retardation times are negligible, the left-propagating output field of the receiving center reads
\begin{align}
\begin{split}
	\Phi^{\rm out, L}_{n,r}(t) & = \Phi^{\rm in, L}_{n,r}(t) + \sqrt{\frac{\gamma_{n,r}(t)}{2}}c_r(t)e^{i\theta_r(t)}, \\
	& = -\Phi^{\rm out, R}_{n,r}(t)e^{i\phi_r^n} + \sqrt{\frac{\gamma_{n,r}(t)}{2}}c_r(t)e^{i\theta_r(t)}, \\
	& = -\left[ \Phi^{\rm in, R}_{n,r}(t) + \sqrt{\frac{\gamma_{n,r}(t)}{2}}c_r(t)e^{i\theta_r(t)}\right]e^{i\phi_r^n} + \sqrt{\frac{\gamma_{n,r}(t)}{2}}c_r(t)e^{i\theta_r(t)}, \\
	& = -\Phi^{\rm out, R}_{n,e}(t)e^{i(\phi_r^n + \phi_{er}^n)} + \frac{1-e^{i\phi_r^n}}{2}\sqrt{\gamma_r(t)}c_r(t)e^{i\theta_r(t)}.
\end{split}
\end{align}
Here, we have considered that the SiV center is equally coupled to both modes ($\beta^n_r =0.5$) so that $\gamma_{n,r}(t) = \gamma_r(t)/2$.

For simplicity, we consider an idealized case of an infinite waveguide where all the reflected signal $\Phi^{\rm out, L}_{n,r}$ never reaches back the emitting center. In that case, the output field of the emitter simplifies to ($\beta^n_e =0.5$)
\begin{align}
\begin{split}
	\Phi^{\rm out, R}_{n,e}(t) & = \Phi^{\rm in, R}_{n,e}(t) + \sqrt{\frac{\gamma_{n,e}(t)}{2}}c_e(t)e^{i\theta_e(t)}, \\
	& = -\Phi^{\rm out, L}_{n,e}(t)e^{i\phi_e^n} + \sqrt{\frac{\gamma_{n,e}(t)}{2}}c_e(t)e^{i\theta_e(t)}, \\
	& = \frac{1 - e^{i\phi_e^n}}{2}\sqrt{\gamma_e(t)}c_e(t)e^{i\theta_e(t)}, \\
	& \equiv \frac{1 - e^{i\phi_e^n}}{2}\Phi(t).
\end{split}
\end{align}
For a perfectly fulfilled dark-state condition $|\Phi^{\rm out, L}_{t}| = 0$, i.e.
\begin{equation}
	\sqrt{\gamma_r(t)}c_r(t)e^{i\theta_r(t)} = \frac{\sin(\phi^t_e/2)}{\sin(\phi^t_r/2)}\Phi(t)e^{i\phi_L^t/2}
	\quad \therefore \quad
	\phi_L^t = \phi_r^t + \phi_e^t + 2\phi_{er}^t,
\end{equation}
the left-propagating longitudinal signal becomes
\begin{align}
	r_l = \left\vert\frac{\Phi^{\rm out, L}_{l,r}(t)}{\Phi(t)\sin(\phi_e^l/2)}\right\vert^2 
	& = \left\vert 1 - \frac{\sin(\phi_e^t/2)}{\sin(\phi_e^l/2)}\frac{\sin(\phi_r^l/2)}{\sin(\phi_r^t/2)}e^{i(\phi^t_L - \phi_L^l)/2}\right\vert^2,  \label{Eq:OutPutL}\\
	& = 1 + \frac{\sin^2(\phi_e^t/2)}{\sin^2(\phi_e^l/2)}\frac{\sin^2(\phi_r^l/2)}{\sin^2(\phi_r^t/2)}
	 - 2\frac{\sin(\phi_e^t/2)}{\sin(\phi_e^l/2)}\frac{\sin(\phi_r^l/2)}{\sin(\phi_r^t/2)}\cos[(\phi^t_L - \phi_L^l)/2]. \nonumber
\end{align}
This results indicate how much signal is emitted in the longitudinal branch when a perfect suppression of the transverse wave occurs. In the infinite waveguide limit, this signal is completely lost and gives a good estimation of the state transfer fidelity. 

One can distinguish two phenomena contributing to the emitting signal. There is the intra-band interference which determines the effective emission rate of each centers into the difference mode, $\tilde \gamma_{j,n} = 2\gamma_{j,n}\sin^2(\phi^n_j/2)$, and is captured by the second term of Eq.~\eqref{Eq:OutPutL}.
Finally, there is the inter-band interference responsible for the third term. It roughly indicates how efficient the driving on the emitting center is to also suppress the emission in the longitudinal branch.

\begin{figure}
\centering
\includegraphics[width= 0.8 \columnwidth]{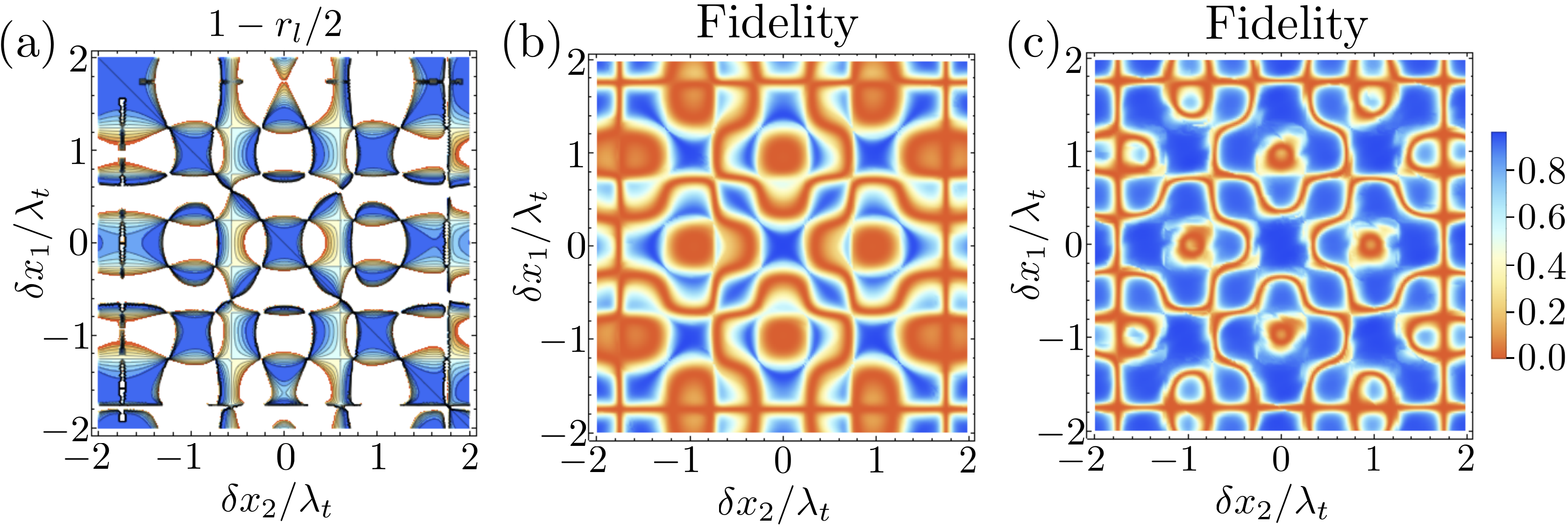}
\caption{State transfer fidelity with time-varying driving as a function of the position of the centers. (a) Fidelity estimation from the left-propagating longitudinal output field from the receiving center, as described by $r_l$ in Eq.~\eqref{Eq:OutPutL}.
(b) Full simulation in the case of an infinite waveguide where all left-propagating emitted field by the receiving center is lost. 
(c) Simulation in the case of a 1mm waveguide ($\Delta\omega_t = 14$).
The center position $\delta x_e = \delta x_r = 0$ corresponds to $\phi_e^n = \phi_r^n = \pi$ and we chose $\phi^t_L - \phi_L^l = 0$.
The other parameters used are as in Fig.~3 (e) of the main text. 
}
\label{Fig:DSCond}
\end{figure}

In Fig.~\ref{Fig:DSCond}, we show the robustness of the state transfer protocol for variations in the positioning of the emitting ($\delta x_e$) and receiving ($\delta x_r$) SiV centers for $\phi^t_L - \phi_L^l = 0$, where $\delta x_r = \delta x_e = 0$ corresponds to maximal couplings $\phi^n_e = \phi^n_r = \pi$. We consider the case of an infinite waveguide where Eq.~\eqref{Eq:OutPutL} gives the proper intuition. Already at this level, the fidelity is robust for small variations, as predicted by a small displacement expansion 
\begin{equation}
	r_l \approx \frac{(k_l - k_t)^2}{4}(\delta x^2_e + \delta x^2_r)^2.
\end{equation}

Finally, we compare to the finite waveguide case, where emitted field in the longitudinal mode can be reabsorbed after round trips within the waveguide. In that case, the protocol becomes more robust and we recover the results shown in Fig.~3 of the main text.

